\documentclass[journal,twocolumn,twoside]{IEEEtran}

\usepackage{graphics} 
\usepackage{graphicx}
\usepackage{subfigure}
\usepackage{amssymb}  
\usepackage{cite}
\usepackage{amsmath}
\usepackage[boxed,ruled,lined]{algorithm2e}
\usepackage{epstopdf}
\usepackage{booktabs}
\usepackage{multirow}
\usepackage{etoolbox}
\usepackage{mathdots}
\usepackage{xcolor, float}
\makeatletter
\patchcmd{\@makecaption}
  {\scshape}
  {}
  {}
  {}
\makeatother
\usepackage{diagbox}
\usepackage{bm}
\usepackage{subfigure}
\usepackage{lipsum}
\usepackage{braket,amsfonts,amsopn}
\usepackage{hyperref}

\def\norm #1{\left\|#1\right\|}

\def\twon #1{\left\|#1\right\|_2}

\def\frobn #1{\left\|#1\right\|_{\text{F}}}

\def\abs #1{\left|#1\right|}
\def\inp #1{\left\langle#1\right\rangle}

\def\st{\text{subject to }}

\def\bC{\mathbb{C}}

\def\bR{\mathbb{R}}

\def\bS{\mathbb{S}}

\def\m #1{\boldsymbol{#1}}

\def\cH{\mathcal{H}}

\def\cL{\mathcal{L}}
\def\cM{\mathcal{M}}

\def\cO{\mathcal{O}}
\def\cP{\mathcal{P}}

\def\cS{\mathcal{S}}
\def\cT{\mathcal{T}}

\def\bee{\begin{equation}}
\def\ene{\end{equation}}

\def\beq{\begin{eqnarray}}
\def\enq{\end{eqnarray}}
\def\lentwo{\setlength\arraycolsep{2pt}}

\newtheorem{lem}{Lemma}

\newtheorem{thm}{Theorem}

\def\equ #1{\begin{equation}#1\end{equation}}
\def\equa #1{\begin{eqnarray}#1\end{eqnarray}}
\def\sbra #1{\left(#1\right)}
\def\mbra #1{\left[#1\right]}
\def\lbra #1{\left\{#1\right\}}
\def\diag #1{\text{diag}#1}
\def\tr #1{\text{tr}#1}

\def\rank #1{\text{rank}#1}
\def\st {\text{ subject to }}

\DeclareMathOperator*{\argmin}{arg\,min}

\title{Multichannel Frequency Estimation in Challenging Scenarios via Structured Matrix Embedding and Recovery (StruMER)}

\author{Xunmeng Wu, Zai Yang, and Zongben Xu
	\thanks{
	The research of the project was supported by the National Natural Science Foundation of China under Grant 61977053. 
	
	The authors are with the School of Mathematics and Statistics, Xi'an Jiaotong University, Xi'an 710049, China (e-mails: wxm1996@stu.xjtu.edu.cn, yangzai@xjtu.edu.cn, zbxu@xjtu.edu.cn). 
	
	Corresponding author: Zai Yang. }
}

\begin{document}
\maketitle

\begin{abstract}
	Multichannel frequency estimation with incomplete data and miscellaneous noises arises in array signal processing, modal analysis, wireless communications, and so on. In this paper, we consider maximum-likelihood(-like) optimization methods for frequency estimation in which proper objective functions are adopted subject to observed data patterns and noise types. We propose a universal signal-domain approach to solve the optimization problems by embedding the noiseless multichannel signal of interest into a series of low-rank positive-semidefinite block matrices of Hankel and Toeplitz submatrices and formulating the original parameter-domain optimization problems as equivalent structured matrix recovery problems. The alternating direction method of multipliers (ADMM) is applied to solve the resulting matrix recovery problems in which both subproblems of ADMM are solved in (nearly) closed form. The proposed approach is termed as \emph{stru}ctured \emph{m}atrix \emph{e}mbedding and \emph{r}ecovery (StruMER). Extensive numerical simulations are provided to demonstrate that StruMER has improved threshold performances in various challenging scenarios, e.g., limited data, low signal-to-noise ratio, impulsive noise, and closely spaced frequencies, as compared with state-of-the-art methods.
\end{abstract}


\section{Introduction}
Multichannel frequency estimation is a fundamental problem in statistical signal processing. It refers to the process of estimating the frequency and amplitude parameters of several complex sinusoids from multichannel signals of their superposition. Multichannel frequency estimation appears in applications such as array signal processing \cite{krim1996two,stoica2005spectral},  modal analysis \cite{park2014modal,li2018atomic}, radar \cite{li2007mimo,zheng2017super}, wireless communications \cite{barbotin2012estimation,garcia2017direct}, and so on \cite{zhang2018multichannel}. As compared to the single-channel case, multichannel signals contain more samples that can be used to improve the estimation performance, but also bring new challenges as to how to exploit the redundancy among the channels. 

Due to its connection to array signal processing, multichannel
frequency estimation has a long history of research; see \cite{krim1996two,stoica2005spectral}. The estimation of frequencies from noisy data is difficult since the multichannel signal is a highly nonlinear function of the frequencies. The maximum likelihood estimation (MLE) method, if solvable, provides benchmark performance for frequency estimation. To realize the MLE, several optimization approaches, e.g., expectation maximization (EM) \cite{feder1988parameter}, Newton-type algorithms \cite{starer1992newton}, the relaxation (RELAX) algorithms \cite{li1996efficient,li1997angle}, and multisnapshot Newtonized orthogonal matching pursuit (MNOMP) \cite{ZHU2019175,7491265}, have been developed to directly solve for the optimal frequencies. But these methods either are heavily sensitive to parameter initialization or have poor performance in the presence of closely spaced frequencies.  

Under the assumption of Gaussian noise and complete data, subspace methods, e.g., the multiple signal classification (MUSIC) \cite{1143830} and the estimation of parameters by rotational invariant techniques (ESPRIT) \cite{roy1989esprit}, and the method of direction estimation (MODE \cite{stoica1990novel,57542}, MODEX \cite{GERSHMAN1999221}), achieve good estimation performance with a sufficient number of channels. But the performance of these methods deteriorates in challenging scenarios in which the number of channels is small, the frequencies are closely spaced, the observed data are incomplete, or the noise is impulsive.

Limited channels and closely spaced frequencies are common in practice. For instance, in MIMO localization \cite{garcia2017direct}, several MIMO base stations are deployed to collect the broadcast signals from the user that are then used to determine the user’s position. In this case, multichannel data are acquired by combing the signals received in multiple base stations in a fusion center. The number of data channels is usually small due to limited number of base stations. The angles of arrival, which are associated with the frequencies of interest, are closely spaced in presence of closely located users.
To achieve good estimation performance in such challenging scenarios, it is of great importance to utilize the full structures of the multichannel signal.

Incomplete data appear in various applications of multichannel frequency estimation \cite{wang2006spectral,stoica2009missing}. In power system monitoring, the operator needs to estimate the state from the synchronized phasor measurements by phasor measurement units (PMU). The incomplete data case arises when some PMU data points do not reach the operator due to PMU malfunction or communication congestions \cite{gao2015missing,zhang2018multichannel}. In array signal processing, an antenna array is deployed to collect the  electromagnetic source signals that are then used to estimate the directions. The incomplete data case arises when some sensors fail to work \cite{zhang2010robust} or sparse linear arrays are adopted \cite{moffet1968minimum}.  
Compressed sensing and atomic norm minimization (ANM) methods have been developed that are flexible in dealing with incomplete data but usually suffer from a resolution limit \cite{malioutov2005sparse,yang2016exact,li2016off,steffens2018compact,duarte2013spectral,tang2013compressed,yang2022separation}.

Impulsive noise is frequently encountered in many practical wireless radio systems. In radar and sonar systems, impulsive noise occurs due to electromagnetic and acoustic interference \cite{zoubir2012robust}. As compared to Gaussian noise, 
the probability density function (PDF) of impulsive
noise has heavier tails, which brings some values several times greater than the standard deviation of Gaussian noise \cite{zeng2013ell}. To perform robust multichannel frequency estimation, many techniques, e.g., fractional lower-order statistics (FLOS) \cite{liu2001subspace}, robust statistics \cite{zoubir2012robust}, $\ell_p$-norm ($1\le p <2$) minimization \cite{zeng2013ell}, and sparse Bayesian learning (SBL) \cite{dai2017sparse,zhou2021robust}, have been exploited to resist impulsive noise. Three representative approaches, including $\ell_p$-MUSIC \cite{zeng2013ell}, Bayes-optimal method \cite{dai2017sparse}, and atom-based Bayesian learning method (BLM) \cite{zhou2021robust}, have achieved good estimation performance. In particular, $\ell_p$-MUSIC consists of two consecutive steps where the $\ell_p$-norm of the residual fitting error matrix is minimized to robustly estimate the noise subspace, from which the frequencies are estimated. Since the spectral structure in multichannel signals is not exploited to solve for the subspace, $\ell_p$-MUSIC is strictly suboptimal. The SBL-type methods \cite{dai2017sparse,zhou2021robust} decompose the impulsive noise into two parts: background Gaussian noise and outliers; and they jointly estimate the unknown sinusoidal parameters and the outliers according to the maximum a posteriori (MAP) criterion. To estimate the continuous-valued frequencies, The Bayes-optimal method uses a root grid refining method which cannot completely resolve the grid mismatch issue \cite{stoica2011sparse} and leads to performance degradation. BLM adopts a dynamic grid model solved by gradient-based methods which require a careful initialization.  

In our recent work \cite{wu2022maximum} on line spectral estimation (a.k.a., single-channel frequency estimation), a structured matrix recovery approach is proposed in which the single-channel signal is embedded into a low-rank positive-semidefinite (PSD) block matrix of Hankel and Toeplitz submatrices. This enables us to solve the maximum-likelihood optimization problem in the signal domain and overcome drawbacks of previous methods that work directly in the parameter domain. It therefore inspires us to develop a similar technique for the multichannel frequency estimation problem concerned in the present paper. But such an extension is challenging in two aspects. First, it was unclear how to embed multichannel signals into structured matrices so that the redundancy among the channels can be exploited and meanwhile the structured matrices can be effectively recovered. Second, the algorithm design is more difficult in presence of incomplete data and impulsive noises. 

In this paper, we propose a structured matrix embedding and recovery (StruMER) algorithm for multichannel frequency estimation that operates in the signal domain, uses full signal structures and is flexible in dealing with incomplete data and impulsive noises. Our main contributions are summarized below.

\begin{itemize}
	\item We consider the maximum-likelihood method for multichannel frequency estimation and extend it to the case of incomplete data and impulsive noises by properly modifying the objective function (see Section \ref{sec:Prob}).
	\item We propose to transform the parameter-domain optimization problems into signal-domain ones by solving equivalently for the noiseless multichannel signal from which the frequencies can be easily retrieved. We further formulate the signal-domain problems as rank-constrained structured matrix recovery ones by embedding the noiseless multichannel signal of interest into a series of low-rank PSD block matrices of Hankel and Toeplitz structured submatrices. We show that the Hankel-Toeplitz block matrices uniquely determine the multichannel signal, and vise versa (see Section \ref{sec:main}).
	\item We present an alternating direction method of multipliers (ADMM) algorithm \cite{boyd2011distributed} for the structured matrix recovery problems, in which both subproblems of ADMM are solved in (nearly) closed form, and analyze its convergence property. We present a dimensionality reduction technique to reduce the computational complexity in the case when the number of channels is large. We also extend StruMER to the case when the number of frequency components is unknown (see Section \ref{sec:ADMM}).
	\item We provide extensive simulation results to verify that StruMER has improved threshold performances in various challenging scenarios, e.g., limited data channels, incomplete data, low signal-to-noise ratio (SNR), impulsive noise, and closely spaced frequencies, as compared to the existing methods (see Section \ref{sec:sim}).
\end{itemize}

The sets of real and complex numbers are denoted $\bR$ and $\bC$, respectively. For vector $\m{x}$, $\m{x}^T$, $\overline{\m{x}}$, $\m{x}^H$, and $\norm{\m{x}}_2$ denote its transpose, complex conjugate, conjugate transpose, and $\ell_2$-norm, respectively. For matrix $\m{X}$, its transpose, complex conjugate, conjugate transpose, inverse, pseudo-inverse, Frobenius norm, rank, column space, and trace are denoted $\m{X}^T$, $\overline{\m{X}}$, $\m{X}^H$, $\m{X}^{-1}$, $\m{X}^{\dagger}$, $\frobn{\m{X}}$, $\rank\sbra{\m{X}}$, $\text{range}\sbra{\m{X}}$, and $\tr\sbra{\m{X}}$, respectively. The inner product of matrices $\m{X}$ and $\m{Y}$ is defined as $\langle \m{X},\m{Y} \rangle_{\bR} = \Re\lbra{\tr\sbra{\m{X}^H\m{Y}}}$. $\m{X} \succeq \m{0}$ means that $\m{X}$ is Hermitian PSD. The notation $\abs{\cdot}$ denotes the modulus of a scalar or the cardinality of a set. The diagonal matrix with vector $\m{x}$ on the diagonal is denoted $\diag\sbra{\m{x}}$. The $j$th entry of vector $\m{x}$ is $x_j$, the $\sbra{i,j}$th entry of $\m{X}$ is $x_{ij}$, and the $i$-th row (or $j$-th column) of $\m{X}$ is $\m{X}_{i,:}$ (or $\m{X}_{:,j}$). For matrix $\m{X}$, its $\ell_p$-norm $\norm{\cdot}_p$ is defined as $\norm{\m{X}}_p = \sbra{\sum_{i}\sum_{j}\abs{x_{ij}}^p}^{1/p}$ and its $\ell_{2,p}$-norm is defined as $\norm{\m{X}}_{2,p} = \sbra{\sum_i \twon{\m{X}_{i,:}}^p}^{1/p}$. An identity matrix is denoted as $\m{I}$. The notation $\bS_K^+$ denotes the set of PSD matrices of rank no greater than $K$. Denote $\cH$ as a Hankel operator that maps $\m{x} \in \bC^{N}$ to an $n \times n$ Hankel matrix $\cH\m{x}$ with its $\sbra{i,j}$ entry given by $x_{i+j-1}$ where $N=2n-1$. Denote $\cT$ as a Hermitian Toeplitz operator that maps proper $\m{t}\in \bC^n$ to an $n\times n$ Hermitian Toeplitz matrix $\cT\m{t}$ with its $\sbra{i,j},i\ge j$ entry given by $t_{i-j+1}$. 

\section{Problem Formulation} \label{sec:Prob}
Let $N\times L$ matrix $\m{X} = \sbra{x_{jl}}$ denote an $L$-channel signal that is composed of $K$ spectral components and is given by:
\equ{
	x_{jl} = \sum^K_{k=1} e^{i2\pi f_k \sbra{j-1}} s_{kl}, \; j = 1,\ldots,N, \; l = 1,\ldots,L, \label{eq:1}
}
where $N$ is the signal length per channel, $i=\sqrt{-1}$, $f_k \in [-1/2,1/2)$ and $s_{kl}$ denote the $k$-th unknown frequency and the associated amplitude in the $l$-th channel, respectively. It is seen that the signals among the multiple channels share the same frequency profile but have different amplitudes. We call $\m{X}$ spectral-sparse since $K$ is usually small. Letting $\m{a}(f)=\mbra{1,e^{i2\pi f},\ldots,e^{i2\pi f\sbra{N-1}} }^T$ represent a complex sinusoid of frequency $f$, $\m{A}(\m{f}) =\mbra{\m{a}(f_1),\ldots,\m{a}(f_K)}$ be a Vandermonde matrix, and $\m{S}$ be the $K \times L$ amplitude matrix formed by $\lbra{s_{kl}}$, we rewrite \eqref{eq:1} as
\equ{
	 \m{X} = \sum^K_{k=1} \m{a}\sbra{f_k} \m{S}_{k,:} = \m{A}(\m{f}) \m{S}, \label{eq:2}
}
where $\m{S}_{k,:}$ is the $k$-th row of $\m{S}$. 

Suppose that we have access to the noisy samples given by:
\equ{
	\m{Y} = \cP_{\Omega} \sbra{ \m{X} + \m{E}}, \label{eq:model}
}
where the index set $\Omega \subseteq \lbra{1,\ldots,N}\times \lbra{1,\ldots,L}$, $\cP_{\Omega}$ is the projection onto the entries supported on $\Omega$ that sets all entries outside of $\Omega$ to zero, and $\m{E}=\sbra{\varepsilon_{jl}}$ is the matrix of additive noise. The objective of frequency estimation is to estimate the frequencies $\lbra{f_k}$ given the multichannel data $\m{Y}$.

In this paper we will pay special attention to some challenging scenarios in which existing methods have difficulties. In particular, we consider the following scenarios which are clarified according to the observed data patterns, complete versus incomplete data, and the noise distribution, Gaussian versus non-Gaussian noise:
\begin{enumerate}
	\item Complete data with Gaussian noise: In this case, $\Omega = \lbra{1,\ldots,N}\times \lbra{1,\ldots,L}$ and $\varepsilon_{jl} \sim \mathcal{CN}\sbra{0,\sigma^2}$. We focus on the scenarios in which the number of channels $L$ is small, or the SNR is low, or the frequencies $\lbra{f_k}$ are closely spaced. 
	\item Incomplete data with Gaussian noise: In this case, $\Omega \subset \lbra{1,\ldots,N}\times \lbra{1,\ldots,L}$ and $\varepsilon_{jl} \sim \mathcal{CN}\sbra{0,\sigma^2}$. We consider two data patterns in which some data are missing either element-wise or row-wise randomly.
	\item Non-Gaussian noise: In this case, $\varepsilon_{jl} \nsim \mathcal{CN}\sbra{0,\sigma^2}$. We consider impulsive noise that occurs at arbitrary locations or in several rows.
\end{enumerate}

In this paper, we propose the following parameter-domain optimization problem to solve for the frequencies:
\equ{
	\min_{\m{f},\m{S}}g\sbra{\m{A}\sbra{\m{f}}\m{S} - \m{Y} }, \label{eq:prb}
}
where $g\sbra{\cdot}$ is the objective function chosen according to the noise distribution and the observed data patterns. For Gaussian noise, we choose $g\sbra{\cdot} = \frobn{\cdot}^2$ in the complete data case and $g\sbra{\cdot} = \frobn{\cP_{\Omega}\sbra{\cdot}}^2$ in the incomplete data case, which makes \eqref{eq:prb} serve as a maximum likelihood estimator. For impulsive noise, we choose $g\sbra{\cdot} = \norm{\cdot}_p^p$, $1\leq p < 2$ in the complete data case and use $g\sbra{\cdot} = \norm{\cP_{\Omega}\sbra{\cdot}}_p^p$, $1\leq p < 2$ in the incomplete data case. The $\ell_p$ norm is adopted because it is less sensitive to outliers than the Frobenius norm and is shown to have good performance in resisting impulsive noise \cite{zeng2013ell,wen2016robust}. For row-impulsive noise, we choose $g\sbra{\cdot} = \norm{\cdot}^p_{2,p}$ and $g\sbra{\cdot} = \norm{\cP_{\Omega}\sbra{\cdot}}^p_{2,p},1\le p < 2$ for complete and incomplete data, respectively.

\section{Structured Matrix Embedding} \label{sec:main}
\subsection{Signal-Domain Optimization Problem}
The parameter-domain optimization problem in \eqref{eq:prb} is highly nonlinear with respect to the frequencies $\lbra{f_k}$ and is difficult to solve directly. In this paper, we assume that the model order $K$ is known unless otherwise stated and reformulate \eqref{eq:prb} as the following signal-domain optimization problem:
\equ{
	\min_{\m{X}}g\sbra{\m{X} - \m{Y}}, \st \m{X}\in S_0, \label{eq:prb1}
}
where
\equ{
	S_0 = \lbra{ \m{A}\sbra{\m{f}}\m{S}: \; f_k \in [-1/2,1/2), \m{S}\in \bC^{K\times L} }
}
denotes the set of multichannel spectral-sparse signals. By \eqref{eq:prb1}, we have moved the nonconvexity from the objective to the constraint. To make \eqref{eq:prb1} a tractable optimization problem, the key is to construct an equivalent and computable surrogate set for $S_0$.

\subsection{Previous Results on Characterization of $S_0$}
\subsubsection{Toeplitz Model} \label{sec:T}
An equivalent reformulation of $S_0$ has been proposed in the literature on ANM methods for frequency estimation when $L=1$ \cite{tang2013compressed} and $L\ge 1$ \cite{yang2016exact,li2016off}.
The following theorem is summarized from \cite{yang2016exact}.

\begin{thm} \label{thm:ST}
	Assume $K<N$. The following statements are true:
\begin{enumerate}
	\item $S_0 = \lbra{\m{X}: \begin{bmatrix}\m{Z} & \m{X}^H \\ \m{X} & \cT \m{t}' \end{bmatrix} \in \bS_K^+ \text{ for some } \m{Z} \text{ and }  \m{t}' }$; 
	\item For any $\m{X} = \m{A}\sbra{\m{f}}\m{S} \in S_0$ with distinct $\lbra{f_k}$ and $p_k = \frac{\twon{\m{S}_{k,:}}}{\sqrt{L}}> 0$, $\begin{bmatrix}\m{Z} & \m{X}^H \\ \m{X} & \cT \m{t}' \end{bmatrix} \in \bS_K^+$ if and only if
    \lentwo{\equa{\m{Z}
    &=& \m{S}^H \sbra{\diag\sbra{\m{p}'}}^{-1} \m{S}, \label{eq:Z} \\ 
    \cT\m{t}'
    &=& \m{A}\sbra{\m{f}}\diag\sbra{\m{p}'}\m{A}^H\sbra{\m{f}}, \label{eq:T1}
    }}where $\m{p}'$ is an arbitrary vector of positive values.
\end{enumerate}
\end{thm}

Using Theorem \ref{thm:ST}, the problem in \eqref{eq:prb1} is written equivalently as the following rank-constrained Toeplitz matrix recovery problem:
\equ{
	\min_{\m{X},\m{Z},\m{t}'} g\sbra{\m{X} - \m{Y}}, \st \begin{bmatrix}\m{Z} & \m{X}^H \\ \m{X} & \cT \m{t}' \end{bmatrix} \in \bS_K^+. \label{eq:prb_T}
}
Suppose that $\m{X}\in S_0$ is an optimal solution to \eqref{eq:prb1}. Then, any $\lbra{\m{Z},\m{t}'}$ satisfying \eqref{eq:Z} and \eqref{eq:T1} with $\m{p}'$ being free variables results in an optimal solution to \eqref{eq:prb_T}. Therefore, there exist infinitely many optimal solutions to \eqref{eq:prb_T} that can even be unbounded. This results in convergence issues for any algorithm used to solve \eqref{eq:prb_T}, which has been partly verified in \cite{wu2022maximum} in the case $L=1$ and will be verified in Section \ref{sec:sim}.

\subsubsection{Structured Matrix Embedding for Single-channel Signals}
The issues of nonuniqueness and unboundedness in the previous Toeplitz model are resolved in \cite{wu2022maximum} in the  case $L=1$. Let 
\equ{
	S_0^1=\lbra{\m{A}\sbra{\m{f}}\m{s}:\; f_k \in [-1/2,1/2), \m{s}\in \bC^K}
} 
denote $S_0$ in this case (i.e., $\m{S}$ in $S_0$ degenerates into $\m{s}$ in $S_0^1$ when $L=1$). We assume that the length $N$ of $\m{X}$ is odd with $N=2n-1$ for integer $n$. If $N$ is even, then we may assume that the $(N + 1)$-st sample is missing, which can be tackled as in the incomplete data case. The following theorem is given in \cite[Theorem 1]{wu2022maximum}.
\begin{thm} \label{thm:SHT1}
	Assume $K<n$. The following statements are true:
	\begin{enumerate}
		\item $S_0^1 = \lbra{\m{x}: \begin{bmatrix} \cT\overline{\m{t}} & \cH \overline{\m{x}} \\ \cH \m{x} & \cT \m{t} \end{bmatrix} \in \bS_K^+ \text{ for some } \m{t} }$; 
		\item For any $\m{x} = \m{A}\sbra{\m{f}}\m{s} \in S^1_0$ with distinct $\lbra{f_k}$ and nonzero $s_k$'s, $\begin{bmatrix} \cT\overline{\m{t}} & \cH \overline{\m{x}} \\ \cH \m{x} & \cT \m{t} \end{bmatrix} \in \bS_K^+$ if and only if
		\equ{
			\cT\m{t} = \m{A}_n\sbra{\m{f}}\diag\sbra{\abs{\m{s}}}\m{A}_n^H\sbra{\m{f}},
		}
		where $\abs{\cdot}$ operates on $\m{s}$ element-wise and $\m{A}_n\sbra{\m{f}} = \mbra{\m{a}_n(f_1),\ldots,\m{a}_n(f_K)}$ is an $n\times K$ Vandermonde matrix with $\m{a}_n(f_k) = \mbra{1,e^{i2\pi f_k},\ldots,e^{i2\pi f_k (n-1)}}^T$ .
	\end{enumerate}
\end{thm}

It is shown in Theorem \ref{thm:SHT1} that any single-channel spectral-sparse signal in $S_0^1$ can be embedded into a low-rank PSD block matrix of Toeplitz and Hankel submatrices, and this embedding is one-to-one. 

Such an embedding technique is extended in \cite{wu2022Direction} to the special direction-of-arrival (DOA) estimation problem with constant modulus (CM) source signals, which corresponds to the case where each row of the amplitude matrix $\m{S}$ must have a CM in \eqref{eq:2} in the language of frequency estimation. Differently from \cite{wu2022Direction}, we consider the typical case with general amplitude matrix $\m{S}$. Therefore, the extension in \cite{wu2022Direction} is not applicable to the multichannel frequency estimation problem concerned in the present paper. It will also been seen that the matrix embedding technique proposed in the ensuing subsection is more mathematically sophisticated as compared to \cite{wu2022Direction}.

\subsection{Proposed Structured Matrix Embedding for Multichannel Signals} \label{sec:HT}
In this subsection, we propose a structured matrix embedding technique for multichannel spectral-sparse signals. Formally, we have the following theorem, which generalizes Theorem \ref{thm:SHT1} from the single-channel to the multichannel case.
\begin{thm} Assume $K < n$. The following statements are true:
\begin{enumerate}
\item \equ{
	\begin{split}
		S_0 = & \left\{ \m{X}:\; \begin{bmatrix}\cT\overline{\m{t}^l} & \cH \overline{\m{X}}_{:,l} \\ \cH \m{X}_{:,l} & \cT \m{t} \end{bmatrix} \in \bS_K^+, \; l=1,\ldots,L \right.\\
		& \phantom{ \ldots,L } \text{ for some } \lbra{\m{t}^l},\m{t} \text{ satisfying } \sum_{l=1}^L\m{t}^l = L\m{t} \bigg\},
	\end{split} \label{eq:part1}
}
where $\cH \m{X}_{:,l} \in \bC^{n\times n}$, $\cT\m{t}^l \in \bC^{n\times n}$,  and $\cT\m{t} \in \bC^{n\times n}$;
\item For any $\m{X} = \m{A}\sbra{\m{f}}\m{S} \in S_0$ with distinct $\lbra{f_k}$ and $p_k = \frac{\twon{\m{S}_{k,:}}}{\sqrt{L}}> 0, \; k=1,\dots,K$, the associated $\lbra{\m{t}^l}$ and $\m{t}$ in \eqref{eq:part1} are uniquely given by
    \lentwo{
    	\equa{
    		\cT\m{t} &=& \m{A}_n\sbra{\m{f}}\diag\sbra{\m{p}}\m{A}_n^H\sbra{\m{f}}, \label{eq:TVandec} \\ 
    		\cT\m{t}^l &=& \m{A}_n\sbra{\m{f}}\diag\sbra{\m{p}^l}\m{A}_n^H\sbra{\m{f}}, \, l=1,\dots,L, \label{eq:TlVandec}
    	}
	}where the $k$-th entry of $\m{p}^l$ is given by
    \equ{
    	p^l_k = \frac{\abs{s_{kl}}^2}{p_k},\quad k=1,\dots,K. \label{eq:pl}
    }
\end{enumerate}
\label{thm:SHTisS0}
\end{thm}

The proof of Theorem \ref{thm:SHTisS0} will be deferred to the subsequent subsection. It is shown in Theorem \ref{thm:SHTisS0}-1) that any multichannel spectral-sparse signal in $S_0$ can be embedded into a number $L$ of low-rank PSD block Hankel-Toeplitz matrices. In the $l$-th block matrix, the Hankel submatrix $\cH\m{X}_{:,l}$ is formed by using the $l$-th channel of the spectral-sparse signal $\m{X}$ of interest. A common Toeplitz submatrix $\cT\m{t}$ is shared among the block matrices to exploit the joint sparsity among the channels. To control the magnitude of the Toeplitz submatrix, a linear constraint is added regarding the distinct Toeplitz submatrices $\lbra{\cT\m{t}^l}$ and the shared Toeplitz submatrix $\cT\m{t}$. Consequently, all the Toeplitz submatrices become bounded and unique under mild conditions. It is seen from \eqref{eq:TVandec} and \eqref{eq:TlVandec} that $\cT\m{t}$ captures the global amplitude information of $\m{S}$, reflected by $\lbra{p_k}$, while each $\cT\m{t}^l$ captures the local amplitude information given by $\lbra{p_k^l}$.

By making use of Theorem \ref{thm:SHTisS0}, the problem in \eqref{eq:prb1} is written equivalently as the following rank-constrained structured matrix recovery problem:
\equ{
	\begin{split}
		& \min_{\m{X},\lbra{\m{t}^l},\m{t}}g\sbra{\m{X} - \m{Y}}, \\
		& \st \begin{bmatrix}\cT\overline{\m{t}^l} & \cH \overline{\m{X}}_{:,l} \\ \cH \m{X}_{:,l} & \cT \m{t} \end{bmatrix} \in \bS_K^+, \, l = 1,\ldots,L, \, \sum_{l=1}^L\m{t}^l = L\m{t}. \label{eq:prb2}
	\end{split}
}
Suppose that the optimal solution to the original parameter-domain problem in \eqref{eq:prb} is given by $\sbra{\m{f}^*, \m{S}^*}$, where it is easy to show that $\m{S}^*_{k,:} \neq \m{0},\; k=1,\ldots,K$ almost surely in the presence of noise. It then follows from Theorem \ref{thm:SHTisS0} that the optimal solution to \eqref{eq:prb2} is given by
\equ{ 
	\begin{split}
		\m{X}^* &= \m{A}\sbra{\m{f}^*} \m{S}^*, \\
		\cT\m{t}^* &= \m{A}_n\sbra{\m{f}^*}\diag\sbra{\m{p}^*}\m{A}_n^H\sbra{\m{f}^*},  \\ 
		\cT\m{t}^{*,l} &= \m{A}_n\sbra{\m{f}^*}\diag\sbra{\m{p}^{*,l}}\m{A}_n^H\sbra{\m{f}^*}, \, l=1,\dots,L, 
	\end{split} \label{eq:Vand}
}
where $p^*_k = {\twon{\m{S}^*_{k,:}}}/{\sqrt{L}}, \; p^{*,l}_k = {\abs{s^*_{kl}}^2}/{p^*_k},\; k=1,\dots,K$. Evidently, $\m{t}^*$ and $\lbra{\m{t}^{*,l}}$ are uniquely determined by $\m{X}^*$. Once the problem in \eqref{eq:prb2} is solved, the frequencies $\m{f}$ can be extracted from $\cT\m{t}$ by computing its Vandermonde decomposition in \eqref{eq:Vand}.

The matrix optimization problem in \eqref{eq:prb2} has certain connections to the ANM methods \cite{li2016off}. Generally speaking, ANM can be viewed as a convex relaxation of the problem in \eqref{eq:prb}, while we solve \eqref{eq:prb} in this paper by the reformulation in \eqref{eq:prb2}. Besides, the problem in \eqref{eq:prb_T} can be viewed as an un-relaxed version of the semidefinite program by which ANM is numerically solved. We show in this paper that \eqref{eq:prb_T} cannot be practically solved, which motivates and is resolved by our structured matrix embedding technique and the problem in \eqref{eq:prb2}.

\subsection{Proof of Theorem \ref{thm:SHTisS0}}
The classical Carath\'{e}odory-Fej\'{e}r's Theorem for Toeplitz matrices \cite[Theorem 11.5]{yang2018sparse} and a lemma regarding factorization of Hankel matrices \cite[Lemma 4]{yang2016vandermonde} play key roles in our proof, which are stated in the following two lemmas.
\begin{lem} \label{lem:T}
	Any PSD Toeplitz matrix $\cT\m{u}\in \bC^{n\times n}$ of rank $K<n$ admits the unique Vandermonde decomposition $\cT\m{u} = \m{A}_n(\m{f}) \diag\sbra{\m{d}} \m{A}^H_n(\m{f})$ where $d_k>0,k=1,\ldots,K$ and $\lbra{f_k}$ are distinct.
\end{lem}

\begin{lem} \label{lem:H}
	If a Hankel matrix $\cH\m{z}$ can be factorized as $\cH\m{z} = \m{A}_n(\m{f}) \m{G}\m{A}^T_n(\m{f})$ where $\m{G}\in \bC^{K \times K}$, $K < n$ and $\lbra{f_k}$ are distinct, then $\m{G}$ must be a diagonal matrix.
\end{lem}

Now, we are ready to prove Theorem \ref{thm:SHTisS0}.
\subsubsection{Proof of the First Part} 
For notational simplicity, we write $\m{A}_n(\m{f})$ as $\m{A}_n$ hereafter. For any $\m{X} = \m{A}\sbra{\m{f}}\m{S} \in S_0$, it is easy to show that $\cH\m{X}_{:,l} = \m{A}_n\diag\sbra{\m{S}_{:,l}}\m{A}_n^T$ for each $l$. Letting $\m{t}$ and $\lbra{\m{t}^l}$ be defined as in \eqref{eq:TVandec} and \eqref{eq:TlVandec}, respectively (we redefine $p_k^l=0$ if $p_k=0$), we have
\equ{
	\begin{split}
		& \cT \sum^L_{l=1} \m{t}^l = \sum^L_{l=1} \cT \m{t}^l = \sum^L_{l=1} \m{A}_n \diag\sbra{\m{p}^l} \m{A}^H_n \\
		& = \m{A}_n \cdot \sum^L_{l=1} \diag\sbra{\m{p}^l} \cdot \m{A}^H_n = \m{A}_n \cdot L  \diag\sbra{\m{p}} \cdot \m{A}^H_n = L  \cT \m{t}, \nonumber
	\end{split}
}
where the first equality follows from the definition of the Hermitian Toeplitz operator $\cT$ and the second, fourth, and last equalities follow from \eqref{eq:TlVandec}, \eqref{eq:pl}, and \eqref{eq:TVandec}, respectively. It follows immediately that $\sum_{l=1}^L\m{t}^l = L\m{t}$. Moreover, it is easy to verify that
\equ{ 
	\begin{split}
		& \begin{bmatrix}\cT\overline{\m{t}^l} & \cH \overline{\m{X}}_{:,l} \\ \cH \m{X}_{:,l} & \cT \m{t} \end{bmatrix} \\
		& = \begin{bmatrix}\overline{\m{A}_n} & \\ & {\m{A}}_n \end{bmatrix} \begin{bmatrix} \diag\sbra{\m{p}^l} & \diag\sbra{\overline{\m{S}}_{:,l}} \\ \diag\sbra{\m{S}_{:,l}} & \diag\sbra{\m{p}}\end{bmatrix} \begin{bmatrix}\overline{\m{A}_n} & \\ & {\m{A}}_n \end{bmatrix}^H.
	\end{split}  \label{eq:decom}
}
In the case when $p_k\neq 0$ for each $k=1,\ldots,K$, we have the factorization
\equ{
	\begin{split}
		& \begin{bmatrix} \diag\sbra{\m{p}^l} & \diag\sbra{\overline{\m{S}}_{:,l}} \\ \diag\sbra{\m{S}_{:,l}} & \diag\sbra{\m{p}}\end{bmatrix} = \begin{bmatrix} \m{I} & \diag\sbra{\overline{\m{S}}_{:,l}} \sbra{\diag\sbra{\m{p}}}^{-1}\\ \m{0} & \m{I} \end{bmatrix} \\
		& \cdot \begin{bmatrix} \diag\sbra{\m{p}^l} - \diag\sbra{\overline{\m{S}}_{:,l}} \sbra{\diag\sbra{\m{p}}}^{-1} \diag\sbra{\m{S}_{:,l}}  & \m{0} \\ \m{0} & \diag\sbra{\m{p}}\end{bmatrix} \\
		& \cdot \begin{bmatrix} \m{I} & \diag\sbra{\overline{\m{S}}_{:,l}} \sbra{\diag\sbra{\m{p}}}^{-1} \\ \m{0} & \m{I} \end{bmatrix}^H,
	\end{split} \label{eq:factor}
}
where the Schur complement of $\diag\sbra{\m{p}}$ satisfies
\equ{
	\diag\sbra{\m{p}^l} - \diag\sbra{\overline{\m{S}}_{:,l}} \sbra{\diag\sbra{\m{p}}}^{-1} \diag\sbra{\m{S}_{:,l}} = \m{0}, \label{eq:Schur}
} 
following from \eqref{eq:pl}. In the case when $p_k=0$ for some $k$, we have $s_{k,l}=0$ and $p^l_k=0$ for each $l=1,\ldots,L$. In this case, \eqref{eq:factor} and \eqref{eq:Schur} still hold by replacing the matrix inverse with the pseudo-inverse. Since $\diag\sbra{\m{p}} \succeq \m{0}$, then \eqref{eq:factor} is PSD and thus \eqref{eq:decom} is PSD for every $l$. Moreover, it follows from \eqref{eq:decom}-\eqref{eq:Schur} that
\equ{
	\begin{split}
		\rank \begin{bmatrix}\cT\overline{\m{t}^l} & \cH \overline{\m{X}}_{:,l} \\ \cH \m{X}_{:,l} & \cT \m{t} \end{bmatrix} 
		&= \rank \begin{bmatrix} \diag\sbra{\m{p}^l} & \diag\sbra{\overline{\m{S}}_{:,l}} \\ \diag\sbra{\m{S}_{:,l}} & \diag\sbra{\m{p}}\end{bmatrix} \\
		& = \rank \sbra{\diag\sbra{\m{p}}} \\
		& \le K.
	\end{split}
}
Consequently, we have $\begin{bmatrix}\cT\overline{\m{t}^l} & \cH \overline{\m{X}}_{:,l} \\ \cH \m{X}_{:,l} & \cT \m{t} \end{bmatrix} \in \bS_K^+, \; l=1,\ldots,L$, and thus $\m{X}$ belongs to the set in \eqref{eq:part1}.

Conversely, for any $\m{X}$ belongs to the set in \eqref{eq:part1}, we need to show that $\m{X}\in S_0$. According to \eqref{eq:part1} and the column inclusion property of PSD matrices \cite[p. 432]{horn2012matrix}, we have
\equ{
	\cT\m{t} \in \bS_K^+,\; \cT\m{t}^l \in \bS_K^+,\; \sum_{l=1}^L \cT\m{t}^l = L \cdot \cT\m{t},\; \cH\m{X}_{:,l} \in \text{range}\sbra{\cT\m{t}}. \label{eq:TTl}
}
Applying Lemma \ref{lem:T}, we have
\equ{
	\cT\m{t} = \m{A}_n\sbra{\widetilde{\m{f}}} \diag\sbra{\widetilde{\m{p}}}\m{A}_n^H\sbra{\widetilde{\m{f}}} \label{eq:TVandec2}
}
for distinct frequencies $\widetilde{\m{f}}$ and vectors $\widetilde{\m{p}}\in \bR^{K}$ of non-negative values.
It then follows from \eqref{eq:TTl} that there exist $K\times n$ matrices $\lbra{\m{B}^l}$ such that $\cH\m{X}_{:,l} = \m{A}_n\sbra{\widetilde{\m{f}}} \m{B}^l = \sbra{\m{B}^l}^T \m{A}^T_n\sbra{\widetilde{\m{f}}}$ where the second equality follows from the symmetry of $\cH\m{X}_{:,l}$. We further have $\m{B}^l= \m{A}^{\dagger}_n\sbra{\widetilde{\m{f}}} \sbra{\m{B}^l}^T \m{A}^T_n\sbra{\widetilde{\m{f}}}$ and thus 
\equ{
	\cH\m{X}_{:,l} = \m{A}_n\sbra{\widetilde{\m{f}}} \m{G}^l \m{A}^T_n\sbra{\widetilde{\m{f}}}, \label{eq:Hxl}
} 
where $ \m{G}^l = \m{A}^{\dagger}_n\sbra{\widetilde{\m{f}}} \sbra{\m{B}^l}^T$ is a $K\times K$ matrix. 
It follows from $K<n$ and Lemma \ref{lem:H} that $\m{G}^l$ must be diagonal, i.e.,
\equ{
	\m{G}^l = \diag\sbra{\widetilde{\m{S}}_{:,l}}  \label{eq:F}
} 
for some matrices $\widetilde{\m{S}}\in \bC^{K\times L}$. It follows immediately from \eqref{eq:Hxl} and \eqref{eq:F} that $\m{X}_{:,l} = \m{A}\sbra{\widetilde{\m{f}}}\widetilde{\m{S}}_{:,l}$, yielding that
\equ{
	\m{X}=\m{A}\sbra{\widetilde{\m{f}}}\widetilde{\m{S}}\in S_0, \label{eq:XX}
}
completing the proof.

\subsubsection{Proof of the Second Part} 
Since $N = 2n-1 \ge 2K + 1$ (note that $n \ge K+1$), we can show that the factorization $\m{X}=\m{A}\sbra{\m{f}}\m{S}$ uniquely determines $\m{f}$ and $\m{S}$ (up to permutations of entries or rows); see, e.g., \cite[Theorem 11.1]{yang2018sparse}. We have also shown that $\m{X}=\m{A}\sbra{\widetilde{\m{f}}}\widetilde{\m{S}}$ in \eqref{eq:XX} for given $\m{X}$ in the set in \eqref{eq:part1}. It then follows from $\m{X} = \m{A}\sbra{\widetilde{\m{f}}}\widetilde{\m{S}} = \m{A}\sbra{\m{f}}\m{S} $ and the unique identifiability that $\widetilde{\m{f}} = \m{f}$ and $\widetilde{\m{S}} = \m{S}$ (up to permutations of entries or rows). 

According to \eqref{eq:TTl} and \eqref{eq:TVandec2}, we have
\equ{
	\cT\m{t}^l = \m{A}_n\diag\sbra{\widetilde{\m{p}}^l}\m{A}_n^H, \quad \sum^L_{l=1} \widetilde{\m{p}}^l = L \widetilde{\m{p}}, \label{eq:Ttl}
}
where $\widetilde{\m{p}}$ and $\widetilde{\m{p}}^l$ are vectors of non-negative values to determine. Then, $\begin{bmatrix}\cT\overline{\m{t}^l} & \cH \overline{\m{X}}_{:,l} \\ \cH \m{X}_{:,l} & \cT \m{t} \end{bmatrix}$ admits the decomposition as in \eqref{eq:decom} and thus
\equ{
	\begin{bmatrix} \diag\sbra{\widetilde{\m{p}}^l} & \diag\sbra{\overline{\m{S}}_{:,l}} \\ \diag\sbra{\m{S}_{:,l}} & \diag\sbra{\widetilde{\m{p}}}\end{bmatrix} \in \bS_K^+. \label{eq:wan}
}
It then follows from \eqref{eq:wan} and \eqref{eq:factor} that the Schur complement of $\diag\sbra{\widetilde{\m{p}}}$ is PSD.
Consequently, for every $k$ and $l$, we have
\equ{
	\widetilde{p}^l_k \widetilde{p}_k \geq \abs{s_{kl}}^2.\label{eq:pklpk}
}
Taking summation over $l$ on both sides of \eqref{eq:pklpk} yields that
\equ{
	 L \widetilde{p}^2_k = \sum_{l=1}^L \widetilde{p}^l_k \widetilde{p}_k \geq \sum_{l=1}^L \abs{s_{kl}}^2 = Lp_k^2>0, \label{eq:p2kgeq}
}
where the first equality follows from \eqref{eq:Ttl} and the second inequality follows from the assumption, and thus $\diag\sbra{\widetilde{\m{p}}}$ has rank $K$. It follows immediately from \eqref{eq:wan} that the Schur complement of $\begin{bmatrix} \diag\sbra{\widetilde{\m{p}}^l} & \diag\sbra{\overline{\m{S}}_{:,l}} \\ \diag\sbra{\m{S}_{:,l}} & \diag\sbra{\widetilde{\m{p}}}\end{bmatrix}$ regarding $\diag\sbra{\widetilde{\m{p}}}$ is zero, or equivalently,
the equality holds always in \eqref{eq:pklpk}, and so is \eqref{eq:p2kgeq}, yielding that
\equ{
	\widetilde{p}_k = \frac{\twon{\m{S}_{k,:}}}{\sqrt{L}}= p_k,\quad \widetilde{p}^l_k = \frac{\abs{s_{kl}}^2}{\widetilde{p}_k} = p^l_k,
}
completing the proof.

\section{Structured Matrix Recovery} \label{sec:ADMM}
While the only nonconvexity of the problem in \eqref{eq:prb2} arises from the $L$ rank constraints, it remains challenging to solve it. Specifically, the gradient of the $\ell_p$-norm $(1\le p<2)$ in the objective function $g\sbra{\cdot}$ is not Lipschitz continuous and conventional proximal gradient algorithms cannot be directly applied. Moreover, the variables $\lbra{\m{t}^l}$ are coupled with $\m{t}$ by a linear constraint so that they cannot be solved separately. 

Note that the ADMM algorithm has shown good convergence and performance for rank-constrained problems \cite{boyd2011distributed, andersson2014new,wu2022maximum}. In this paper, we apply ADMM to solve the structured matrix recovery problem in \eqref{eq:prb2}. To make ADMM work, we introduce auxiliary variables $\lbra{\m{Q}^l \in \bC^{2n\times 2n}}_{l=1}^L$ and define
\equ{ 
	\cS_{\m{t}} = \lbra{\lbra{\m{t}^1,\dots,\m{t}^L,\m{t}}:\; \sum_{l=1}^L\m{t}^l = L\m{t}}. 
}
Using these notations, the problem in \eqref{eq:prb2} is rewritten as
\equ{
	\begin{split}
		& \min_{\lbra{\m{X}, \lbra{\m{t}^1,\dots,\m{t}^L,\m{t}}\in \cS_{\m{t}}}, \lbra{\m{Q}^l\in \bS_K^+}} g\sbra{\m{X} - \m{Y}}, \\
		& \st \m{Q}^l = \begin{bmatrix}\cT\overline{\m{t}^l} & \cH \overline{\m{X}}_{:,l} \\ \cH \m{X}_{:,l} & \cT \m{t} \end{bmatrix}, \; l=1,\ldots,L. \label{eq:prb3}
	\end{split}
}
By \eqref{eq:prb3}, we have divided the optimization variables into two groups that are associated with two subproblems of ADMM, as shown later. In the ADMM framework, we write the augmented Lagrangian function as \cite{boyd2011distributed}:
\equ{
	\begin{split}
		& \cL := \cL\sbra{\m{X}, \lbra{\m{t}^l},\m{t},\lbra{\m{Q}^l},\lbra{\m{\Lambda}^l}} \\
		&= g\sbra{\m{X} - \m{Y}} + \sum_{l=1}^L \inp{\m{Q}^l - \begin{bmatrix}\cT\overline{\m{t}^l} & \cH \overline{\m{X}}_{:,l} \\ \cH \m{X}_{:,l} & \cT \m{t} \end{bmatrix}, \m{\Lambda}^l}_{\bR} \\
		& \quad + \frac{\mu}{2} \sum_{l=1}^L \frobn{\m{Q}^l - \begin{bmatrix}\cT\overline{\m{t}^l} & \cH \overline{\m{X}}_{:,l} \\ \cH \m{X}_{:,l} & \cT \m{t} \end{bmatrix}}^2 \\
		&= g\sbra{\m{X} - \m{Y}} + \frac{\mu}{2}\sum_{l=1}^L\frobn{\m{Q}^l - \begin{bmatrix}\cT\overline{\m{t}^l} & \cH \overline{\m{X}}_{:,l} \\ \cH \m{X}_{:,l} & \cT \m{t} \end{bmatrix} + \frac{1}{\mu}\m{\Lambda}^l}^2 \\
		& \quad - \frac{1}{2\mu}\sum^L_{l=1} \frobn{\m{\Lambda}^l}^2,
	\end{split} 
}
where $\lbra{\m{\Lambda}^l\in \bC^{2n\times 2n}}^L_{l=1}$ compose the Lagrangian multipliers and $\mu > 0$ is the penalty coefficient of the augmented second-order term. The ADMM consists of the iterations:
{
	\lentwo\equa{
		\lbra{\m{Q}^l} &\leftarrow& \argmin_{\lbra{\m{Q}^l} \in \bS_K^+} \cL,  \label{eq:subP1} \\
		\lbra{\m{X},\lbra{\m{t}^l},\m{t}} &\leftarrow& \argmin_{\lbra{\m{X},\lbra{\m{t}^1,\dots,\m{t}^L,\m{t}}\in \cS_{\m{t}}}} \cL, \label{eq:subP2} \\
		\m{\Lambda}^l &\leftarrow& \m{\Lambda}^l + \mu\sbra{\m{Q}^l - \begin{bmatrix}\cT\overline{\m{t}^l} & \cH \overline{\m{X}}_{:,l} \\ \cH \m{X}_{:,l} & \cT \m{t} \end{bmatrix}},  \label{eq:subP3} \\ 
		& & \; l = 1,\ldots,L, \nonumber
	}
}where the latest values of the other variables are always used.

\subsection{Solving the Subproblem in \eqref{eq:subP1}} 
Solving \eqref{eq:subP1} results in the update \cite{dax2014low}:
\equ{
	\m{Q}^l \leftarrow \cP_{\bS_K^+}\sbra{\begin{bmatrix}\cT\overline{\m{t}^l} & \cH \overline{\m{X}}_{:,l} \\ \cH \m{X}_{:,l} & \cT \m{t} \end{bmatrix} - \frac{1}{\mu}\m{\Lambda}^l}, \label{eq:Q_update}
}
for each $l=1,\dots,L$, where the projection $\cP_{\bS_K^+}\sbra{\cdot}$ is obtained as the truncated eigen-decomposition of the matrix argument by setting all but the largest $K$ (or less) positive eigenvalues to zero.

\subsection{Solving the Subproblem in \eqref{eq:subP2}} 
The objective function $\cL$ is separable in $\m{X}$ and $\lbra{\lbra{\m{t}^l},\m{t}}$, and thus they can be solved separately. Denote $\m{W}^l = \m{Q}^l+\frac{1}{\mu}\m{\Lambda}^l$ and write $\m{W}^l= \begin{bmatrix}\m{W}^l_{1} & \sbra{\m{W}^l_{2}}^H \\ \m{W}^l_{2} & \m{W}^l_{3} \end{bmatrix}$ and $\m{\Lambda}^l= \begin{bmatrix}\m{\Lambda}^l_{1} & \sbra{\m{\Lambda}^l_{2}}^H \\ \m{\Lambda}^l_{2} & \m{\Lambda}^l_{3} \end{bmatrix}$ that are partitioned as the Hankel-Toeplitz matrices. 

\subsubsection{Solving for $\mathbf{X}$}
By using the symmetry of Hankel matrices, the problem regarding $\m{X}$ is given by
\equ{
	\begin{split}
		\m{X} \leftarrow \argmin_{\m{X}} g\sbra{\m{X}-\m{Y}} + \mu \sum^L_{l=1} \frobn{\cH\m{X}_{:,l}-\m{W}^l_2}^2.
	\end{split} \label{eq:pX}
}

\paragraph{The Case of Complete Data}
We put the case of Gaussian noise and impulsive noise together for consideration. When $g\sbra{\cdot} = \norm{\cdot}_p^p, 1\le p \le 2$, where the case of $p=2$ corresponds to the Frobenius norm $\frobn{\cdot}^2$, the term $\norm{\m{X} - \m{Y}}^p_p$ is separable in all entries $\lbra{x_{jl}}$ of $\m{X}$. According to the definition of Hankel matrix, the problem regarding $x_{jl}$ is given by
\equ{
	\min_{x_{jl}} \abs{x_{jl} - y_{jl}}^p + \mu \sum_{a+b-1=j} \abs{x_{jl} -  \sbra{\m{W}_2^l}_{ab} }^2,
}
or equivalently,
\equ{
	\min_{x_{jl}} \abs{x_{jl} - y_{jl}}^p + \mu d^\cH_{j} \abs{x_{jl} - \sbra{\cH^H\m{W}_2^l}_j / d^{\cH}_j }^2,
}
where $\cH^H$ denotes the adjoint of the Hankel operator and $\m{d}^\cH = \mbra{1,2,\dots,n,n-1,\dots,1} \in \bR^{N}$.
Denote $\beta_j = \mu d_{j}^{\cH}$. Consequently, the solution (and the update) is given by \cite{combettes2011proximal}
\equ{
	x_{jl}\leftarrow y_{jl} + \text{prox}_{\abs{\cdot}^p,\beta_j}\sbra{ \sbra{\cH^H\m{W}_2^l}_j / d^{\cH}_j - y_{jl} }, \label{eq:X_update}
}
where the proximity operator
\equ{
	\text{prox}_{\abs{\cdot}^p,\beta}\sbra{a} = \text{sign}\sbra{a} z 
}
and $0 \leq z \leq \abs{a}$ is the solution to
\equ{
	2\beta z +  p z^{p-1} - 2 \beta \abs{a} = 0. \label{eq:z}
}
In the case when $p=1$, a solution to \eqref{eq:z} does not exist if $\abs{a}< \frac{1}{2\beta}$, and we have simply $z=0$. This results in the well-known soft-thresholding operator
\equ{
	\text{prox}_{\abs{\cdot}^1,\beta}\sbra{a} = \text{sign}\sbra{a} \max\lbra{\abs{a}-\frac{1}{2\beta},0}.
}
In the case when $p = \frac{5}{4}, \frac{4}{3}, \frac{3}{2}, \frac{5}{3}, \frac{7}{4}, 2$, the solution to \eqref{eq:z} can be obtained by solving a linear, quadratic, cubic or quartic equation that has closed-form solutions.
In other cases, a simple Newton's method can be implemented to solve \eqref{eq:z}, as in \cite{wen2016robust}.

In the case of row-impulsive noise, $\m{X}$ can be similarly updated and details will be deferred to Appendix \ref{append:row_noise}.

\paragraph{The Case of Incomplete Data} Consider $g\sbra{\cdot} = \norm{\cP_{\Omega}\sbra{\cdot}}_p^p$. If $\sbra{j,l} \in \Omega$, we have the update in \eqref{eq:X_update}; otherwise, we have
\equ{
	x_{jl} \leftarrow \sbra{\cH^H\m{W}_2^l}_j / d^{\cH}_j. \label{eq:out}
}
Consider $g\sbra{\cdot} = \norm{\cP_{\Omega}\sbra{\cdot}}_{2,p}^p$. If $\sbra{j,l} \in \Omega$, we have the update in \eqref{eq:row-impulsive_in}; otherwise, we have \eqref{eq:out}.

\subsubsection{Solving for $\lbra{\lbra{\mathbf{t}^l},\mathbf{t}}$}
The problem regarding $\lbra{\lbra{\m{t}^l},\m{t}}$ is given by
\equ{
	\begin{split}
		& \min_{\lbra{\m{t}^l},\m{t}} \sum_{l=1}^L \frobn{\cT\m{t}^l - \overline{\m{W}^l_{1}}}^2 + \frobn{\cT\m{t} - \m{W}^l_{3}}^2, \\
		& \st \sum_{l=1}^L\m{t}^l = L\m{t}. \label{eq:subpt}
	\end{split}
}
We use the method of Lagrange multipliers to solve \eqref{eq:subpt}. In particular, the Lagrangian is given by
\equ{
	\cL' = \sum_{l=1}^L \frobn{\cT\m{t}^l - \overline{\m{W}^l_{1}}}^2 + \frobn{\cT\m{t} - \m{W}^l_{3}}^2 + \inp{\sum_{l=1}^L\m{t}^l - L\m{t}, \m{\lambda}}_{\bR},
}
where $\m{\lambda} \in \bC^{N}$ is the Lagrangian multiplier.
Taking derivatives with respect to the variables, we have
\lentwo{
	\equa{
		\frac{\partial\cL'}{\partial \m{t}^l} &=& 2\cT^H\sbra{\cT\m{t}^l - \overline{\m{W}^l_1}} + \m{\lambda} = \m{0}, \quad l=1,\dots,L, \label{eq:pd1}\\ 
		\frac{\partial\cL'}{\partial \m{t}} &=& 2\sum_{l=1}^L \cT^H\sbra{\cT\m{t} - \m{W}^l_3} - L\m{\lambda} = \m{0}, \label{eq:pd2}\\ 
		\frac{\partial\cL'}{\partial \m{\lambda}} &=& \sum_{l=1}^L\m{t}^l - L\m{t} = \m{0}, \label{eq:pd3}
	}
}where $\cT^H$ denotes the adjoint operator of $\cT$.
Making use of \eqref{eq:pd1} and \eqref{eq:pd3}, we have that
\equ{
	\begin{split}
		\m{0} = \sum_{l=1}^L \frac{\partial\cL'}{\partial \m{t}^l}
		& = 2\cT^H\sbra{\cT\sbra{\sum_{l=1}^L \m{t}^l} - \sum_{l=1}^L \overline{\m{W}^l_1}} + L\m{\lambda} \\
		& = 2\sum_{l=1}^L\cT^H\sbra{\cT\m{t} - \overline{\m{W}^l_1}} + L\m{\lambda},
	\end{split} \label{eq:no_tl}
}
which together with \eqref{eq:pd2} yields that
\equ{
	\m{\lambda} = \frac{1}{L} \sum_{l=1}^L \cT^H\sbra{\overline{\m{W}^l_1} - \m{W}_3^l}. \label{eq:splambda}
}
Making use of \eqref{eq:pd1}, \eqref{eq:splambda}, and \eqref{eq:pd3}, we have the updates
\lentwo{
	\equa{ 
		\m{t}^l &\leftarrow& \sbra{\cT^H\cT}^{-1} \cT^H\sbra{\overline{\m{W}_1^l} - \frac{1}{2L} \sum_{q=1}^L \sbra{\overline{\m{W}_1^q} - \m{W}_3^q}},  \label{eq:updatetl} \\ 
		\m{t} &\leftarrow& \frac{1}{L}\sum_{l=1}^L\m{t}^l. \label{eq:updatet}
	}
}

\begin{algorithm} 	\label{alg:alg_HT}
	\SetAlgoLined
	\caption{\emph{Stru}ctured \emph{m}atrix \emph{e}mbedding and \emph{r}ecovery (StruMER) for multichannel frequency estimation.}
		\KwIn{Observation $\m{Y}$, model order $K$, the objective function $g\sbra{\cdot}$}
		Initialize $\m{X},\lbra{\m{t}^l},\m{t}$, $\lbra{\m{\Lambda}^l}$\;
		\While{not converged}{
		Update $\lbra{\m{Q}^l}$ using \eqref{eq:Q_update}.
		
		\uIf{$g\sbra{\cdot}=\norm{\cdot}^p_p$}{
		 Update $\m{X}$ using \eqref{eq:X_update};
	 	}
		\uElseIf{$g\sbra{\cdot}=\norm{\cdot}^p_{2,p}$}{
		 Update $\m{X}$ using \eqref{eq:row-impulsive};
		}
		\uElseIf{$g\sbra{\cdot}=\norm{\cP_{\Omega}\sbra{\cdot}}^p_p$}{
		 Update $x_{jl}$ using \eqref{eq:X_update} if $\sbra{j,l} \in \Omega$, or using \eqref{eq:out} otherwise\;
		}
		\uElseIf{$g\sbra{\cdot}=\norm{\cP_{\Omega}\sbra{\cdot}}^p_{2,p}$}{
		 Update $x_{jl}$ using \eqref{eq:row-impulsive_in} if $\sbra{j,l} \in \Omega$, or using \eqref{eq:out} otherwise;
		}
		Update $\lbra{\m{t}^l}$ using \eqref{eq:updatetl};
		
		Update $\m{t}$ using \eqref{eq:updatet};
		
		Update $\lbra{\m{\Lambda}^l}$ using \eqref{eq:subP3}.
		}
		Calculate the solution $\sbra{\m{f}^*,\m{S}^*}$ by computing the Vandermonde decomposition of $\cT\m{t}^*$ in \eqref{eq:Vand} using Root-MUSIC \cite{barabell1983improving}\;
		\KwOut{Solution $\sbra{\m{f}^*,\m{S}^*}$ as estimate of $\sbra{\m{f},\m{S}}$}
\end{algorithm}

We summarize the proposed algorithm in Algorithm \ref{alg:alg_HT}, which is named as structured matrix embedding and recovery (StruMER).

\subsection{Convergence Analysis}
The global convergence of ADMM has been extensively studied and understood for convex optimization problems \cite{boyd2011distributed}, while the studies for nonconvex problems are still in progress \cite{xu2012alternating,jiang2014alternating,li2015global,wang2019global}. Inspired by \cite{xu2012alternating,jiang2014alternating}, we provide convergence analysis for StruMER. We denote $\m{z}=\lbra{\m{X},\lbra{\m{t}^l}}$, $g(\m{z})=g(\m{X}-\m{Y})$, and $\cM \m{z} = \lbra{ \begin{bmatrix}\cT\overline{\m{t}^l} & \cH \overline{\m{X}}_{:,l} \\ \cH \m{X}_{:,l} & \frac{1}{L} \sum^L_{q=1} \cT \m{t}^q \end{bmatrix} }^L_{l=1} $ that is linear in (the real and complex parts of) $\m{z}$. Then the problem in \eqref{eq:prb3} can be rewritten as
\equ{
		\min_{\m{z}, \lbra{\m{Q}^l}} g\sbra{\m{z}} + \sum^L_{l=1}\delta_{\bS_K^+}\sbra{\m{Q}^l}, 
		\st \lbra{\m{Q}^l} = \cM\m{z}, \label{eq:P_conver}
}
where $\delta_{\bS_K^+}\sbra{\cdot}$ is the indicator function: $\delta_{\bS_K^+}\sbra{\m{Q}}=0$ if $\m{Q} \in \bS_K^+$ or $\infty$ otherwise. Then we have the following theorem, of which the detailed proof is similar to that of \cite[Theorem 2]{wu2022maximum} and will be omitted.

\begin{thm} \label{thm:Convergence}
	Let $\lbra{\lbra{\m{Q}^l_m}, \m{z}_m, \lbra{\m{\Lambda}^l_m}}$ be a sequence generated by StruMER, where $m$ denotes the iteration index. Assume that
	\equ{
		\lim_{m \to \infty}  \frobn{ \m{z}_{m+1} - \m{z}_m }^2 +
		\sum^L_{l=1} \frobn{ \m{\Lambda}^l_{m+1} - \m{\Lambda}^l_m }^2 = 0. \label{eq:assum}
	}
	Then for any limit point $\lbra{\lbra{\m{Q}^l_*}, \m{z}_*, \lbra{\m{\Lambda}^l_*}}$, $\lbra{\m{z}_*, \lbra{\m{Q}^l_*}}$ is a stationary point of \eqref{eq:P_conver}, i.e.,
	\equ{
		\begin{split}
			& \m{0} \in \partial \delta_{\bS_K^+}\sbra{\m{Q}^l_*} + \m{\Lambda}^l_*, \; l = 1,\ldots,L, \\
			& \nabla g \sbra{\m{z}_*} = \cM^H \lbra{\m{\Lambda}_*^l}, \\
			& \lbra{\m{Q}^l_*} = \cM \m{z}_*,
		\end{split} \label{eq:stationary}
	}
	where $\partial \delta_{\bS_K^+}$ is the general subgradient \cite[Definition 8.3]{rockafellar2009variational} and $\cM^H$ is the adjoint operator of $\cM$.
\end{thm}

Theorem \ref{thm:Convergence} states if the solution sequence produced by StruMER converges, then it converges to a stationary point. Since the mapping $\cM$ is not surjective in our problem, stronger convergence analyses in \cite{li2015global,wang2019global} are not applicable for StruMER. Extensive numerical simulations will be provided in Section \ref{sec:sim} to verify the convergence of StruMER.

\subsection{Computational Complexity} \label{sec:DR}
The computations of StruMER are dominated by the projection $\cP_{\bS_K^+}$ in \eqref{eq:Q_update} that is computed by the truncated eigen-decomposition and thus has a computational complexity of $\cO(N^2K)$ \cite[Section 3.3.2]{halko2011finding}. Consequently, the total computational complexity per iteration is $\cO(N^2KL)$. 

In the case when $L \gg N$ (consider, e.g., DOA estimation), we present a dimensionality reduction technique, as in \cite[Section 11.4.3.3]{yang2018sparse}, to reduce the computational complexity of StruMER to $\cO(N^2K \cdot \min\lbra{N,L})$ when $g\sbra{\cdot}=\frobn{\cdot}^2$ and $g\sbra{\cdot}=\frobn{\cP_{\Omega}\sbra{\cdot}}^2$ where data in some rows are lost in the presence of Gaussian noise. 
Take $g\sbra{\cdot}=\frobn{\cdot}^2$ for example. Let $\m{Y}_{\text{R}} $ be the $N\times N$ matrix $ \sbra{\m{Y}\m{Y}^H}^{1/2} $ (in fact, any matrix $\m{Y}_{\text{R}}$ satisfying that $\m{Y}_{\text{R}}\m{Y}_{\text{R}}^H=\m{Y}\m{Y}^H$). Note that there always exists a unitary matrix $\m{V}$ such that $\m{Y}_{\text{R}}=\m{Y}\m{V}\m{D}_N^T $ where $\m{D}_N = \begin{bmatrix} \m{I}_N & \m{0} \end{bmatrix} \in \bR^{N\times L}$. Suppose that $\sbra{\m{f}^*,\m{S}^*}$ is the solution to the original problem in \eqref{eq:prb}, it can be shown that the solution to the reduced problem $\min_{\m{f},\m{S}_{\text{R}}} \frobn{ \m{A}\sbra{\m{f}}\m{S}_{\text{R}} - \m{Y}_{\text{R}} }^2$ is $\sbra{\m{f}^*,\m{S}^*_{\text{R}}}$ with $\m{S}^*_{\text{R}} = \m{S}^*\m{V}\m{D}^T_N$, which yields the same frequency solution $\m{f}^*$ (see a similar proof in \cite[Section 11.4.3.3]{yang2018sparse}). Then we can formulate an equivalent reduced Hankel-Toeplitz matrix recovery model for the reduced problem by making the substitutions $\m{Y} \rightarrow \m{Y}_{\text{R}}$ and $\m{X} \in \bC^{N\times L} \rightarrow \m{X}_{\text{R}}\in \bC^{N\times N}$ in \eqref{eq:prb2}. The reduced model shares the same frequency solution with \eqref{eq:prb2}, but it has only $N$ (instead of $L$) matrix constraints, which reduces the number of eigen-decompositions in StruMER.

Besides, for any $g\sbra{\cdot}$, the complexity can be reduced to $\cO(N^2K + N^2L)$ by using parallel computing since the updates in \eqref{eq:Q_update} are independent among the channels, where $N^2L$ arises from the updates of $\lbra{\m{X},\lbra{\m{t}^l},\m{t}}$.

\subsection{Extension: Model Order Selection}
We have assumed that the model order $K$ is known in StruMER. When it is unknown, its estimation is known as model order selection, which is an equally important problem. Existing methods are usually based on data covariance matrix such as the predicted eigen-threshold approach \cite{chen1991detection} or information theoretic criteria \cite{stoica2004model}. Since StruMER is based on the deterministic maximum likelihood estimation, we use StruMER jointly with an information-theoretic approach, e.g., Akaike information criterion (AIC) and Bayesian information criterion (BIC), to do model order selection. In particular, in an information-theoretic approach, the model order is estimated as $K^*$, among candidates in $\lbra{1,\dots,\widetilde{K}_{\max}}$, that minimize the criterion:
\equ{
	 K^* = \argmin_{\widetilde{K} \in \lbra{1,\ldots,\widetilde{K}_{\max}}} \lbra{-2\ln p\sbra{\m{Y}|\m{\theta}^*_{\widetilde{K}}} + \eta \cdot n_{\widetilde{K}}}, \label{eq:mos}
}
where $\m{\theta}^*_{\widetilde{K}} = \lbra{\m{f}^*_{\widetilde{K}}, \m{S}^*_{\widetilde{K}},\sigma^*_{\widetilde{K}}}$ is the maximum likelihood estimate for the model with the candidate model order $\widetilde{K}$, $p\sbra{\m{Y}|\m{\theta}^*_{\widetilde{K}}}$ is the likelihood function of $\m{Y}$, $n_{\widetilde{K}}$ is the number of real parameters (i.e. the model complexity), and $\eta$ is the penalty coefficient associated with information criterion. It follows from \cite{stoica2004model} that $n_{\widetilde{K}} =\sbra{2L+1}\widetilde{K}+1$ and $\eta = 2$ for AIC, and $n_{\widetilde{K}} =\sbra{2L+3}\widetilde{K}+1$ and $\eta = \ln \sbra{2NL}$ for BIC. Evidently, the key to accomplishing the task is to solve the maximum likelihood problems associated with the candidate values of $\widetilde{K}$, which we propose to accomplish by using StruMER. 

\section{Numerical Simulations} \label{sec:sim} 
In this section, we perform numerical experiments to illustrate the performance of the proposed StruMER, especially in several challenging scenarios. For StruMER, $\m{X}$ is initialized by the observation $\m{Y}$, $\lbra{\m{\Lambda}^l_1}$ and $\lbra{\m{\Lambda}^l_3}$ are initialized with the complex Gaussian distribution whose variance is chosen to be proportional to the signal magnitude, and the remaining variables are initialized with zero. We initialize the penalty parameter $\mu$ in ADMM by $\mu_0 = 1/\sqrt{NL}$ and adaptively update it as in \cite[Section 3.4.1]{boyd2011distributed} to accelerate convergence. The ADMM will be terminated if the absolute and relative errors are below $10^{-4}$ and $10^{-5}$, respectively (see \cite[Section 3.3.1]{boyd2011distributed} for details), or a maximum number of $3000$ iterations are reached. In our simulations, the amplitudes $\lbra{s_{kl}}$ are independently generated from a complex Gaussian distribution. The SNR is defined as the ratio of the power of $\m{S}$ to noise power. The root mean squared error (RMSE) of the frequency estimates is computed as $\sqrt{\frac{1}{K}\twon{\m{f}^*-\m{f}}^2}$ and then averaged over $100$ Monte Carlo runs.
Note that the ADMM algorithm can also be used to solve the Toeplitz model in \eqref{eq:prb_T} and the details are presented in Appendix \ref{append:T} for completeness. This approach is termed as the Toeplitz method and will be included for comparison.

\begin{figure}[htbp]
	\centerline{\includegraphics[width=8cm]{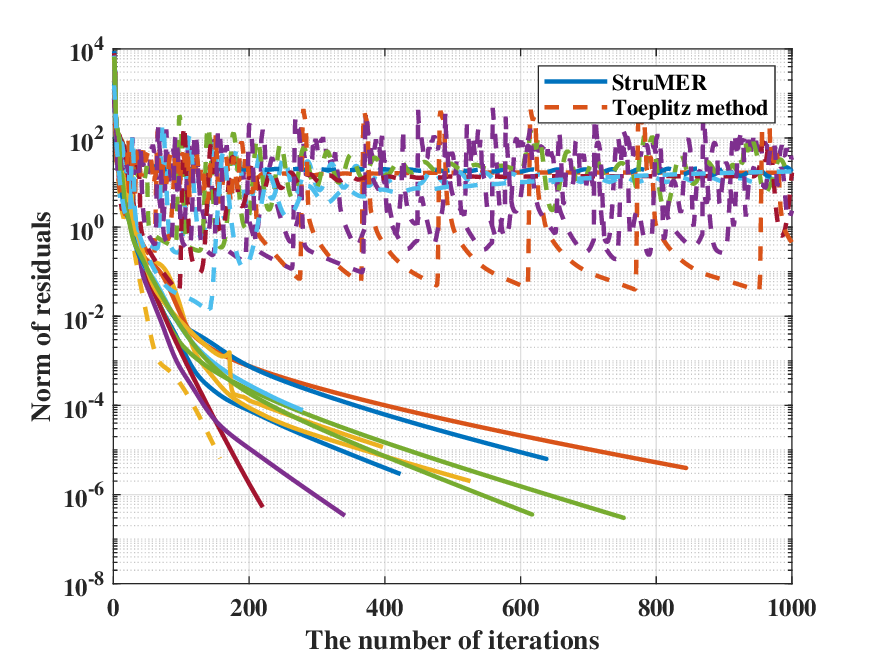}}
	\caption{Norm of residuals versus the number of iterations in $10$ Monte Carlo runs. Solid lines: StruMER; Dotted lines: the Toeplitz method.}
	\label{fig:convergence}
\end{figure}

\subsection{Convergence}
In \emph{Experiment 1}, we evaluate the convergence property of StruMER. We consider a number of $N = 45$ uniform samples with $L=3$ channels that are generated containing white complex Gaussian noise. The SNR is $10$dB and the frequencies are $\m{f} = [-0.2,0.1,0.11]^T$. We conduct StruMER and the Toeplitz method on $10$ randomly generated problems. Fig.~\ref{fig:convergence} presents the norm of residuals in \eqref{eq:assum} versus the number of iterations. It can be seen that as the iteration number increases, the residuals in most trials of the Toeplitz method oscillate around a large value. In contrast to this, all residuals of StruMER tend to converge to zero thanks to uniqueness of its solution. It follows from Theorem \ref{thm:Convergence} that the solution sequence of StruMER converges to a stationary point.

\begin{figure}[htbp]
	\centerline{\includegraphics[width=8cm]{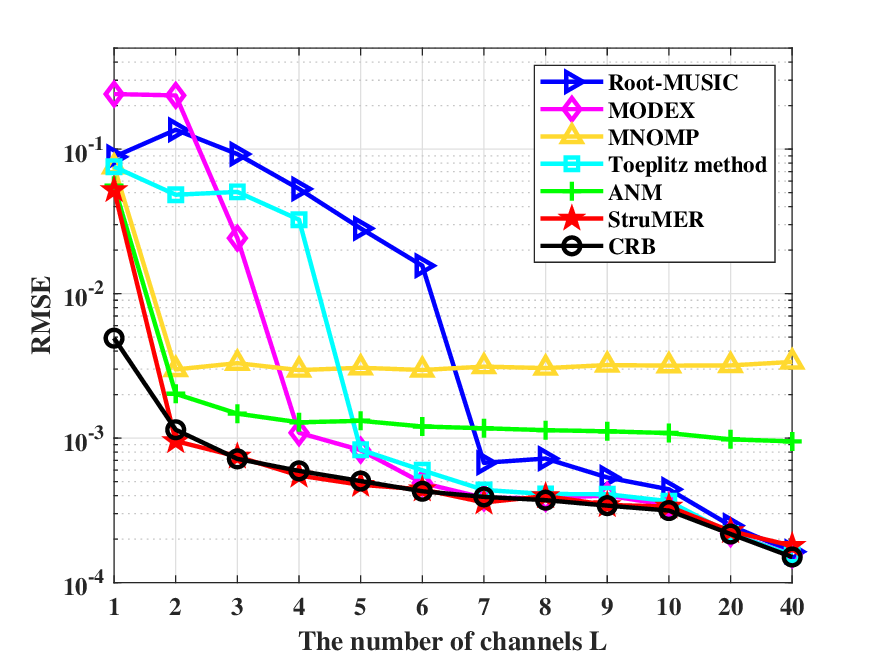}}
	\caption{RMSE versus the number of channels for complete data in Gaussian noise.}
	\label{fig:ULA_L}
\end{figure}

\subsection{The Case of Limited Channels and Closely Spaced Frequencies}
In \emph{Experiment 2}, we study the RMSE performance with respect to the number of channels $L$. The methods that we use for comparison include root-MUSIC \cite{barabell1983improving}, MODEX \cite{GERSHMAN1999221}, MNOMP \cite{ZHU2019175}, and ANM \cite{li2016off}. In MODEX, we use $p=4$ extra-roots and stochastic maximum likelihood function to select the final estimates; see the cited paper for details. We feed into MNOMP the true model order $K$ for a fair comparison and set the other parameters as suggested in \cite{ZHU2019175}. In ANM, the true noise variance is used to calculate the regularization parameter and the true model order $K$ is used to compute the frequency estimates from the Toeplitz matrix for more stable performance. The Cram\'{e}r-Rao bound (CRB) \cite{stoica1989music} is presented as a benchmark that provides a lower bound for any unbiased estimator. 

In the following simulations, we set $N=45$, $\m{f} = [-0.2,0.1,0.11]^T$, and SNR $=10$dB with Gaussian noise unless otherwise specified. Our simulation results are presented in Fig.~\ref{fig:ULA_L}. It is seen that the RMSE of StruMER decreases with the increase of $L$ and attains or is slightly better than the CRB when $L\ge 2$ (note that the maximum likelihood estimator is generally biased and can possibly be better than the CRB \cite{mardia1999bias}). StruMER outperforms the Toeplitz method whose performance suffers from convergence issues. As compared to StruMER, the conventional Root-MUSIC and MODEX require more channels to match the CRB. MNOMP always produces an error greater than the CRB in this case. ANM performs better than MNOMP but there is also a clear gap between ANM and the CRB since it is a convex relaxation method.

\begin{figure}[htb]  
	\centering
	\subfigure[]
	{\includegraphics[width=8cm]{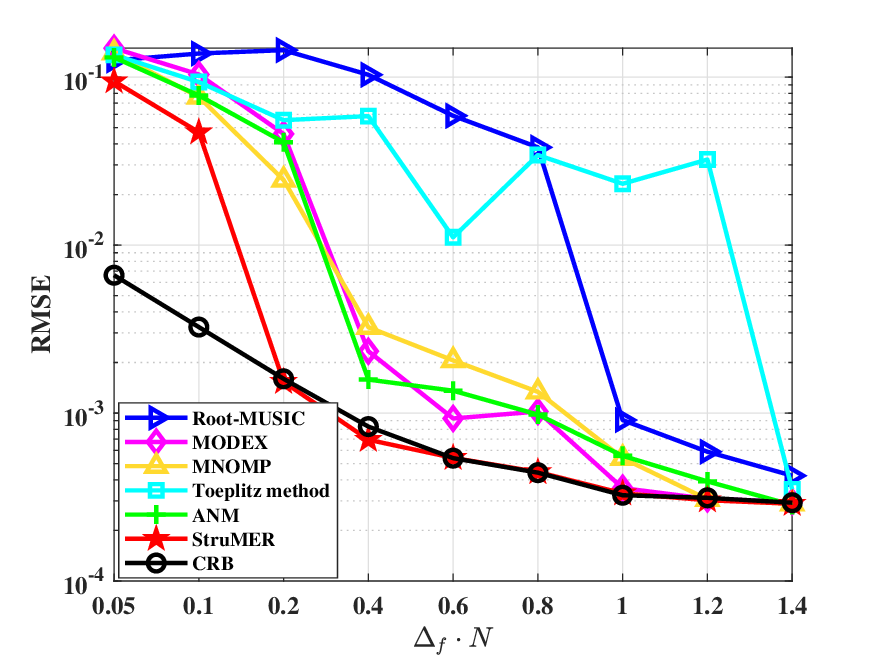}}
	\subfigure[]
	{\includegraphics[width=8cm]{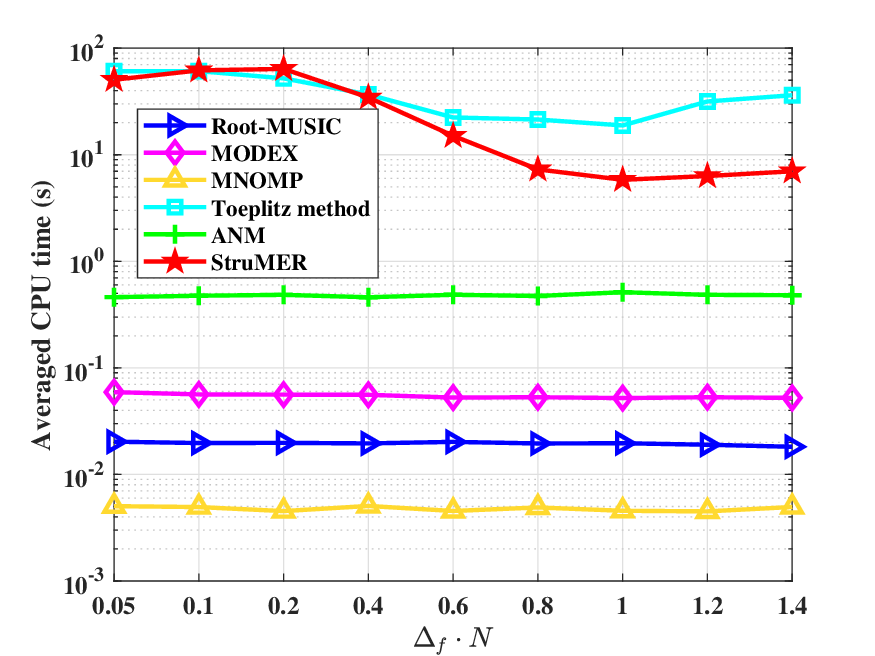}}
	\caption{(a) RMSE and (b) CPU time versus the frequency separation for complete data in Gaussian noise.}
	\label{fig:delta_f}
\end{figure}

In \emph{Experiment 3}, we study the effect of frequency separation $\Delta_f$ on the estimation performance. The number of channels is $L=3$. We consider $\m{f} = [-0.2,0.1,0.1+\Delta_f]^T$ and vary $\Delta_f \in \lbra{0.05,0.1,\ldots,1.4}/N$. Our simulation results are presented in Fig.~\ref{fig:delta_f}. It is seen from Fig.~\ref{fig:delta_f}(a) that StruMER achieves the highest resolution. It is seen from Fig.~\ref{fig:delta_f}(b) that as the frequency separation increases (i.e., when the problem becomes easier), the CPU time of StruMER decreases since fewer iterations are required to converge. Thanks to its good convergence, StruMER is faster than the Toeplitz method, but it is slower than the other methods, mainly due to the computations of the truncated eigen-decompositions in each iteration.

\begin{figure}[htbp]
	\centerline{\includegraphics[width=8cm]{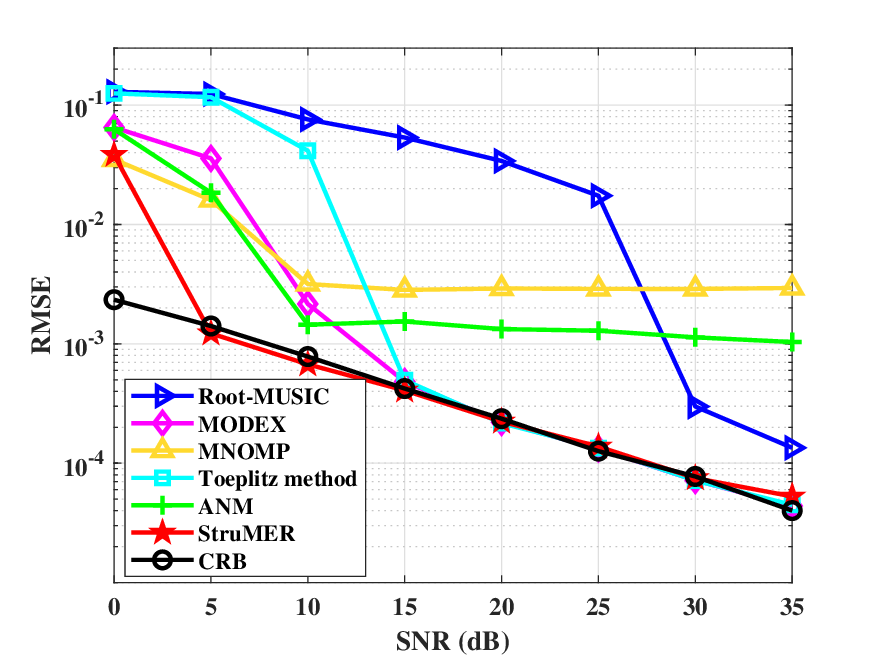}}
	\caption{RMSE versus SNR for complete data in Gaussian noise.}
	\label{fig:ULA_SNR}
\end{figure}

In \emph{Experiment 4}, we study the RMSE performance versus SNR. We set 
$L=3$ and vary the SNR from $0$ to $35$ dB. It is seen from Fig.~\ref{fig:ULA_SNR} that StruMER attains the CRB when SNR $\ge 5$dB. MODEX and the Toeplitz method achieve the CRB only when SNR $\ge15$dB. Root-MUSIC does not attain the CRB due to the limited channels. It is also seen that the performances of MNOMP and ANM slightly improve when SNR $\ge10$dB in this case with closely spaced frequencies.

\subsection{The Case of Incomplete Data}
\begin{figure}[htb]  
	\centering
	\subfigure[] {\includegraphics[width=4.37cm]{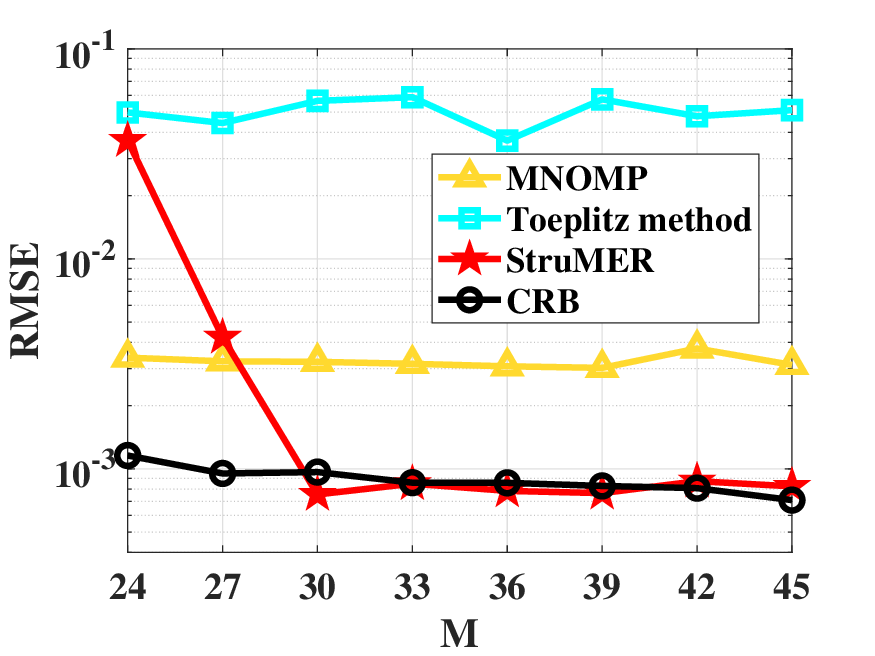}}
	\subfigure[]
	{\includegraphics[width=4.37cm]{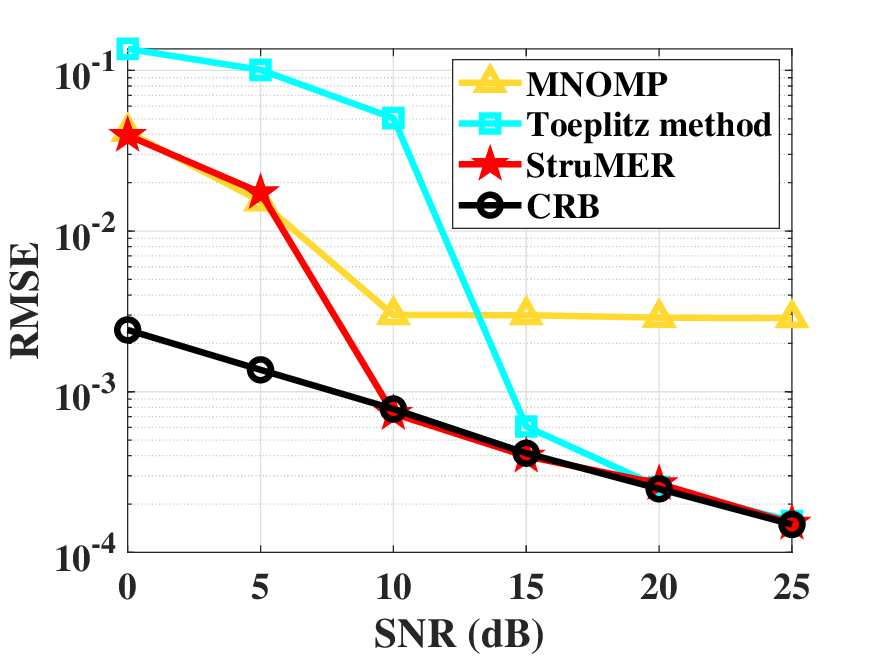}}
	\caption{Mode 1 in the incomplete data case. (a) RMSE versus the number $M$ of observed rows when SNR $=10$dB. (b) RMSE versus SNR when $M=40$.}
	\label{fig:IncompleteData_SLA}
\end{figure}

\begin{figure}[htb]  
	\centering
	\subfigure[]
	{\includegraphics[width=4.37cm]{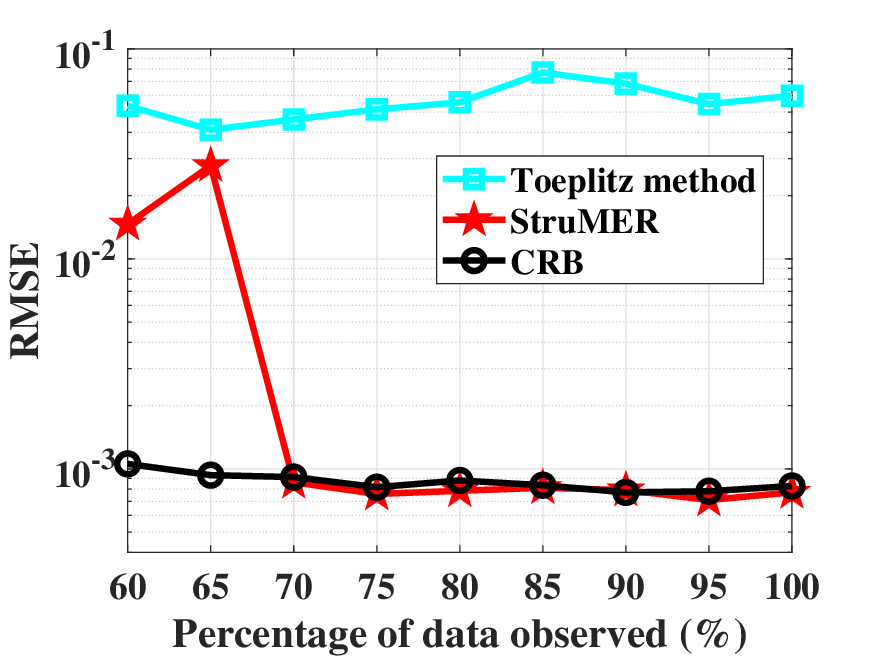}}
	\subfigure[]
	{\includegraphics[width=4.37cm]{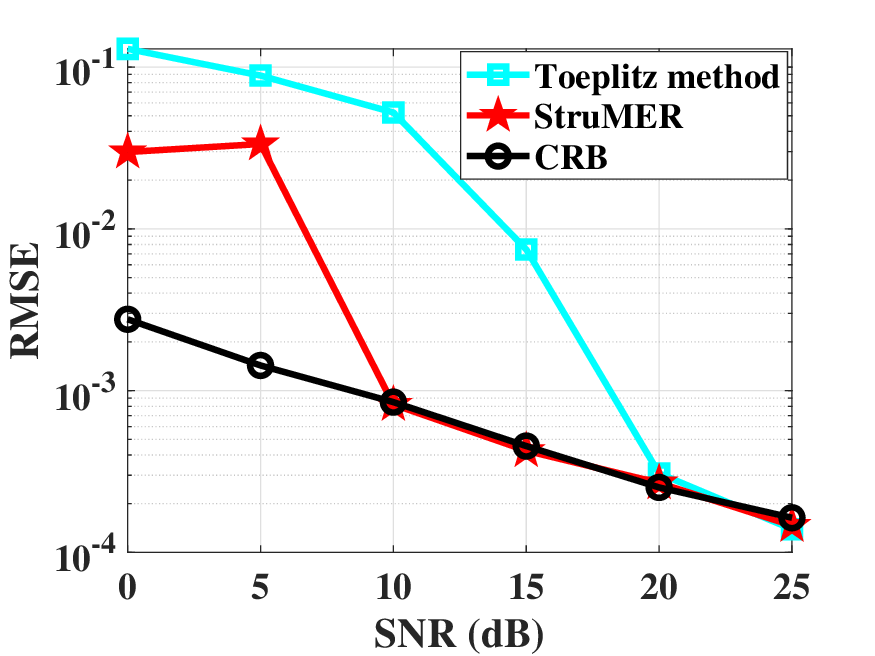}}
	\caption{Mode 2 in the incomplete data case. (a) RMSE versus percentage of data observed when SNR $=10$dB. (b) RMSE versus SNR when $80\%$ data are observed.}
	\label{fig:IncompleteData_random}
\end{figure}

In \emph{Experiment 5}, we consider the incomplete data case where some data are missing randomly in a row-wise (Mode 1) or element-wise (Mode 2) pattern. We set $N=45$, $L=3$, $\m{f}=\mbra{-0.2,0.1,0.11}^T$. Following \cite{stoica1989music}, the CRB of frequency estimation in the incomplete data case is given by
\equ{
	\begin{split}
		& \text{CRB}^{-1}\sbra{\m{f}}  = \\
		& \frac{2}{\sigma^2} \sum^L_{l=1} \Re \lbra{ \m{B}^H_l \m{D}^H \mbra{ \widetilde{\m{\Omega}}_l - \widetilde{\m{\Omega}}_l \m{A}\sbra{\m{A}^H\widetilde{\m{\Omega}}_l\m{A}}^{-1} \!\! \m{A}^H \widetilde{\m{\Omega}}_l } \m{D}\m{B}_l },
	\end{split} \nonumber
}
where $\m{B}_l = \diag\sbra{\m{S}_{:,l}}$, $\m{D}=\mbra{d\m{a}(f_1)/df_1,\ldots,d\m{a}(f_K)/df_K}$, and $\widetilde{\m{\Omega}}_l = \diag\sbra{\Omega_{:,l}}$. Root-MUSIC and MODEX are not applicable in this case. In Mode 1, we plot the RMSE versus the number $M$ of observed rows and the SNR in Fig.~\ref{fig:IncompleteData_SLA}(a) and (b), respectively. In Mode 2, similar results are plotted in Fig.~\ref{fig:IncompleteData_random} where MNOMP is omitted since it cannot be directly applied in this case. It is seen that StruMER performs the best among all the methods. The Toeplitz method has comparable performance as StruMER only in the high SNR regime. As in the complete data case, MNOMP cannot achieve the CRB.

\subsection{The Case of Impulsive Noise}
As in \cite{zeng2013ell,dai2017sparse}, we use the Gaussian mixture model (GMM) to generate the impulsive noise $\varepsilon_{jl}$. The PDF of the two-term GMM is 
\equ{
	p_n(\varepsilon) = \sum^2_{i=1} \frac{c_i}{\pi\sigma^2_i} \exp\sbra{-\frac{\abs{\varepsilon}^2}{\sigma_i^2}},
}
where $\sigma_i^2$ and $0\le c_i \le 1$ are the variance and  probability
of the $i$-th term, respectively, with $c_1 + c_2 = 1$. We set $\sigma^2_2=100\sigma^2_1$ and $c_2=0.1$. The SNR here is redefined as the ratio of the power of $\m{S}$ to $\sigma_1^2$. 
The methods that we use for comparison include the $\ell_p$-MUSIC estimator \cite{zeng2013ell}, the Bayes-optimal method \cite{dai2017sparse}, and the CRB for GMM noise \cite{kozick2000maximum}. As suggested in \cite{zeng2013ell,dai2017sparse}, we take $p=1.1$ for $\ell_p$-MUSIC and the grid interval $1^\circ$ in the Bayes-optimal method. The Toeplitz method is omitted in the following simulations due to its poor performance.

\begin{figure}[htbp]
	\centerline{\includegraphics[width=8cm]{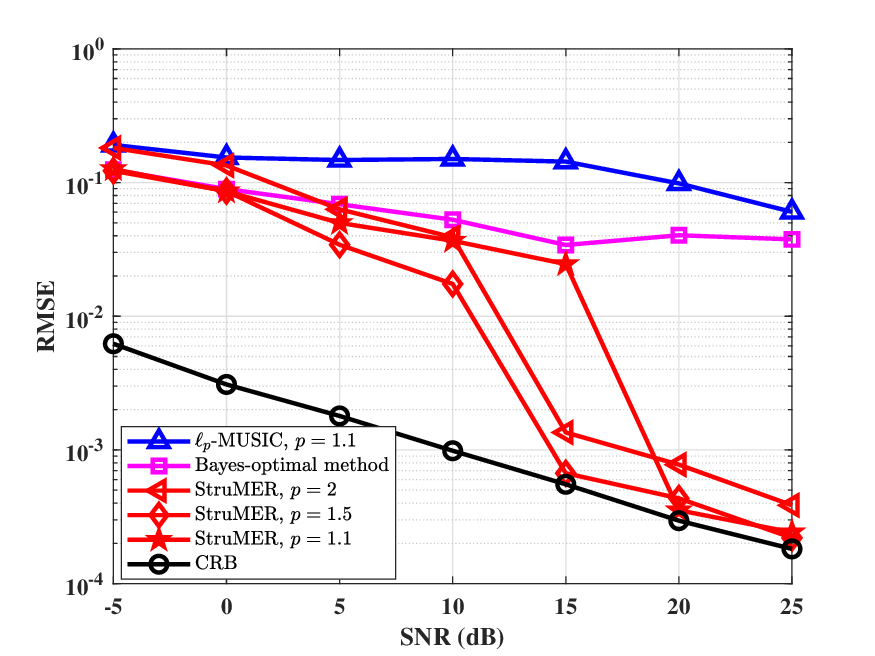}}
	\caption{RMSE versus SNR for complete data in impulsive noise with frequencies $\m{f}=\mbra{-0.2,0.1,0.11}^T$.}
	\label{fig:Impulsive_near}
\end{figure}

\begin{figure}[htb]  
	\centering
	\subfigure[] {\includegraphics[width=8cm]{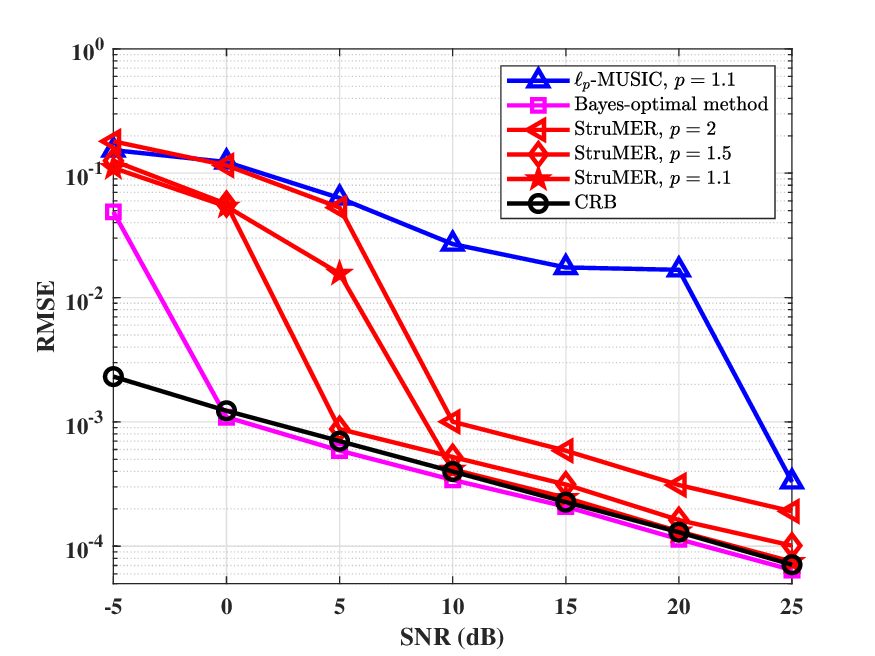}}
	\subfigure[]
	{\includegraphics[width=8cm]{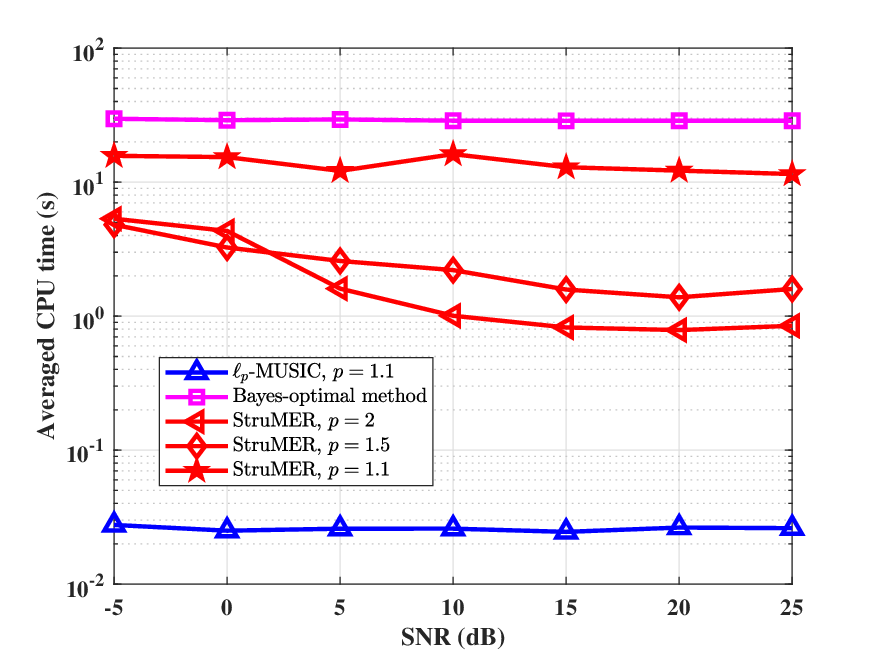}}
	\caption{(a) RMSE and (b) CPU time versus SNR for complete data in impulsive noise with frequencies $\m{f}=\mbra{-0.2,0.1,0.13}^T$.}
	\label{fig:Impulsive_far}
\end{figure}

In \emph{Experiment 6}, we set $N=45$ and $L=3$ and consider $p\in \lbra{1.1,1.5,2}$ for StruMER. The results with $\m{f}=\mbra{-0.2,0.1,0.11}^T$ and $\m{f}=\mbra{-0.2,0.1,0.13}^T$ are presented in Fig.~\ref{fig:Impulsive_near} and Fig.~\ref{fig:Impulsive_far}, respectively. It is seen from Fig.~\ref{fig:Impulsive_near} that thanks to the full use of multichannel signal structures, StruMER outperforms $\ell_p$-MUSIC and the Bayes-optimal method in the case of closely spaced frequencies. When the frequencies are well-separated in Fig.~\ref{fig:Impulsive_far}(a), the performance of StruMER is better than $\ell_p$-MUSIC and is slightly inferior to the Bayes-optimal method. It is seen from Fig.~\ref{fig:Impulsive_far}(b) that with the decrease of $p$, StruMER takes more time since more iterations are required to converge and there is no closed-form solution to \eqref{eq:z} when $p=1.1$. StruMER is slower than $\ell_p$-MUSIC but faster than the Bayes-optimal method whose computational complexity per iteration is $\cO\sbra{N(G+N)^2L}$ where $G$ is the grid number.

\subsection{Combination of Aforementioned Scenarios}
\begin{figure}[htb]  
	\centering
	\subfigure[] {\includegraphics[width=4.37cm]{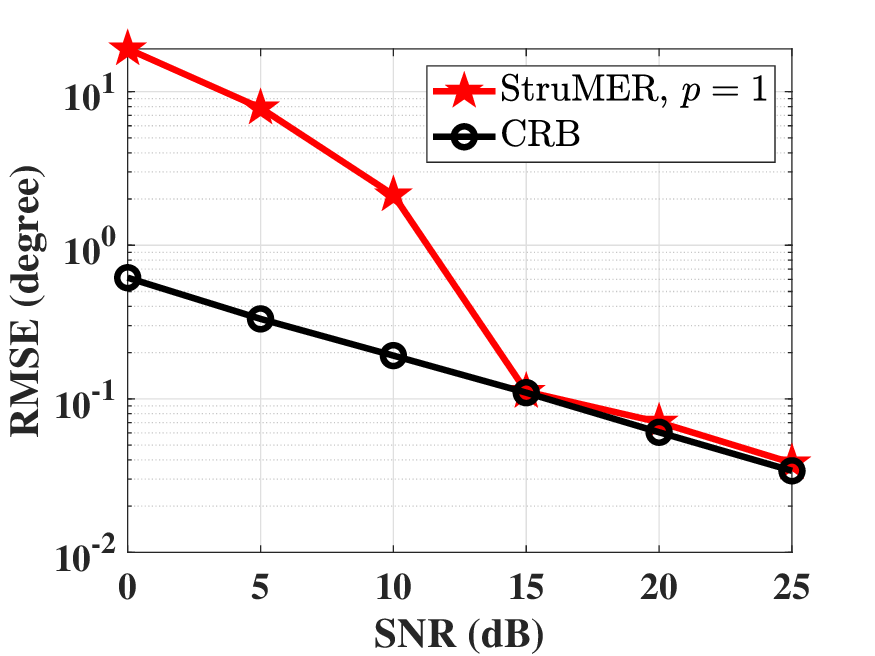}}
	\subfigure[]
	{\includegraphics[width=4.37cm]{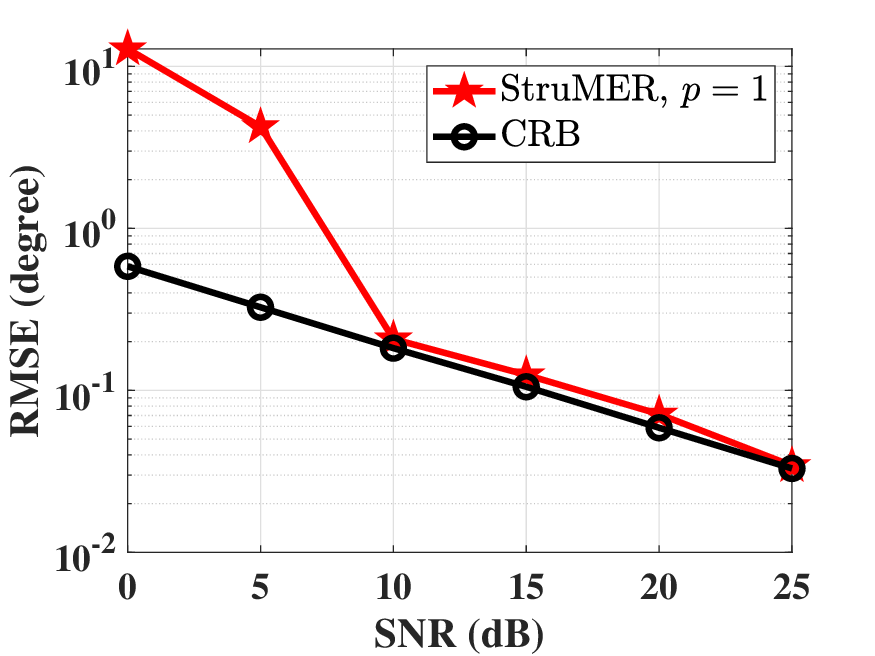}}
	\caption{RMSE (degree) versus SNR in DOA estimation for impulsive noise environments when (a) $80\%$ data are observed and (b) a SLA with $M=13$ is used.}
	\label{fig:Impulsive_M}
\end{figure}

In \emph{Experiment 7}, we study the performance of StruMER in the application to DOA estimation where a combination of the aforementioned scenarios is considered. Assume that $K$ far-field narrowband sources impinge on a uniform linear array (ULA) in which adjacent sensors are placed by half a
wavelength apart. The ULA is equipped with $N = 15$ sensors and the DOAs are $\m{\theta} = \mbra{-1^\circ,5^\circ,40^\circ}^T$ where the first two sources are closely located. The number of snapshots is $L=10$. The source signals are assumed to follow the complex Gaussian distribution. The measurements are contaminated with impulsive noise as in \emph{Experiment 6}. We consider two incomplete data cases where $20\%$ data randomly loss or a sparse linear array (SLA) composed of $M=13$ sensors is used. The aforementioned methods for comparison are not applicable in this case. Our simulation results are presented in Fig.~\ref{fig:Impulsive_M}. It is seen that StruMER with $p=1$ achieves the CRB in the regime of moderate SNR, validating the superiority of StruMER in such a challenging scenario.

\subsection{Dimensionality Reduction}
\begin{figure}[htbp]
	\centerline{\includegraphics[width=8cm]{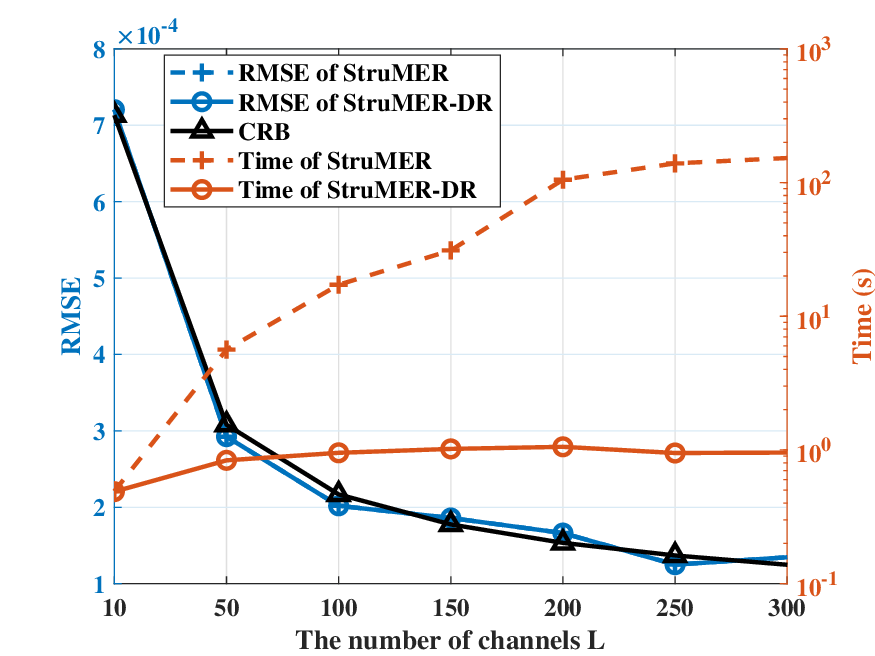}}
	\caption{RMSE and computational time of StruMER and StruMER-DR versus the number of channels.}
	\label{fig:DR}
\end{figure}

In \emph{Experiment 8}, we evaluate the performance of StruMER equipped with the dimensionality reduction technique in Section \ref{sec:DR}, which is termed as StruMER-DR. We set $N = 15$, SNR $=10$dB, and $\m{f} = [-0.2,0.1,0.3]^T$. The results are presented in Fig.~\ref{fig:DR}. It is seen that StruMER-DR shares nearly the same RMSE as StruMER that achieves the CRB. Moreover, the computational time of StruMER-DR keeps almost constant and is much less than that of StruMER as $L$ increases from $50$ to $300$.

\subsection{Model Order Selection}
\begin{figure}[htbp]
	\centerline{\includegraphics[width=8cm]{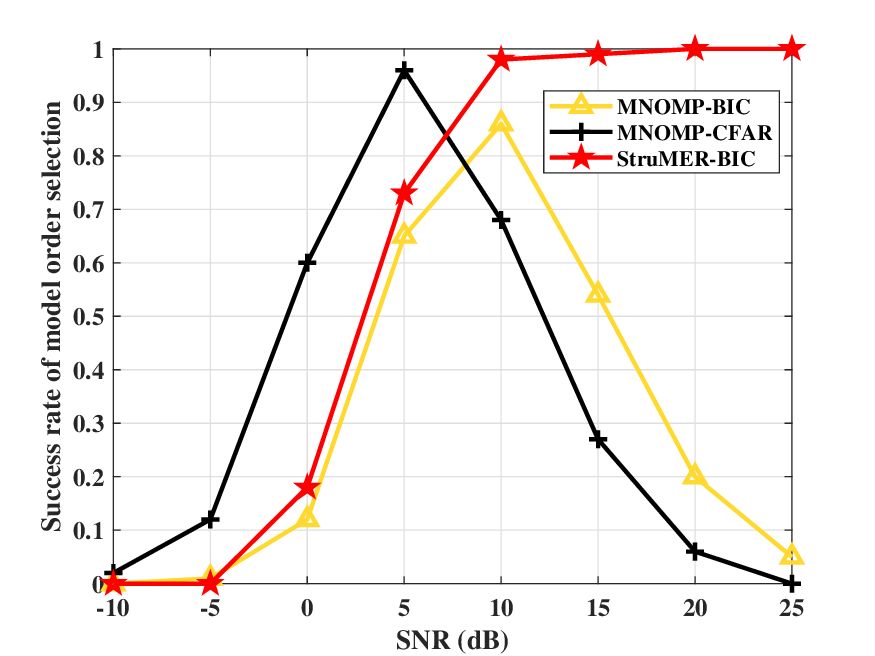}}
	\caption{Results of model order selection.}
	\label{fig:Model_order}
\end{figure}

In \emph{Experiment 9}, we study the model order selection performance by combing StruMER and the BIC criterion, referred to as StruMER-BIC. We repeat the simulation settings in \emph{Experiment 4}. MNOMP with BIC or the constant false alarm rate (CFAR) criterion \cite{ZHU2019175} is added for comparison. The false alarm rate is fixed to $10^{-2}$.
Without loss of generality, the maximum possible model order is set as $\widetilde{K}_{\max}=5$ in \eqref{eq:mos}. For each SNR, $100$ Monte Carlo runs are carried out to compute the success rate. It is seen from Fig.~\ref{fig:Model_order} that thanks to the accurate frequency estimates, StruMER-BIC always outperforms MNOMP-BIC. The success rate of StruMER-BIC approaches one as the SNR increases. In contrast to this, the success rates of MNOMP-BIC and MNOMP-CFAR are peaked around SNR $= 10$dB and SNR $= 5$dB, respectively. This is because MNOMP has poor performance in the presence of closely spaced frequencies and results in overestimated model order.

\section{Conclusion}
In this paper, the StruMER approach was proposed for multichannel frequency estimation by embedding the multichannel spectral-sparse signal into a series of low-rank PSD Hankel-Toeplitz structured matrices and presenting an ADMM algorithm to solve the resulting matrix recovery problem. The proposed approach has great flexibility in dealing with the scenarios of limited channels, incomplete data, and impulsive noise. Extensive numerical simulations were provided to demonstrate the superiority of the proposed approach in the above challenging scenarios as well as in the low-SNR or high-resolution regime.

The computational speed of StruMER is restricted by truncated eigen-decompositions in each iteration of ADMM. It is interesting to apply existing techniques to accelerate the truncated eigen-decompositions (see, e.g., \cite{halko2011finding,liu2013limited}) and develop fast nonconvex algorithms (see, e.g., \cite{absil2008optimization}) in the future.

\appendix
\subsection{Solving \eqref{eq:pX} in the Case of Row-Impulsive Noise} \label{append:row_noise}
When $g\sbra{\cdot}=\norm{\cdot}^p_{2,p}$, the term $\norm{\m{X} - \m{Y}}^p_{2,p}$ is separable in all rows $\lbra{\m{X}_{j,:}}$ of $\m{X}$. The problem regarding $\m{X}_{j,:}$ is given by
\equ{
	\min_{\m{X}_{j,:}} \twon{\m{X}_{j,:}-\m{Y}_{j,:}}^p + \mu d^{\cH}_j \twon{\m{X}_{j,:}- \m{b}^j / d^{\cH}_j }^2, \label{eq:row_noise}
}
where $\m{b}^j=\mbra{\sbra{\cH^H\m{W}_2^1}_j,\ldots,\sbra{\cH^H\m{W}_2^L}_j} \in \bC^{1\times L}$. By letting $\m{Z}_{j,:}=\m{X}_{j,:}-\m{Y}_{j,:}$, \eqref{eq:row_noise} is rewritten as
\equ{
	\min_{\m{Z}_{j,:}} \twon{\m{Z}_{j,:}}^p + \mu d^{\cH}_j \twon{\m{Z}_{j,:}- \sbra{ \m{b}^j / d^{\cH}_j - \m{Y}_{j,:}} }^2.
}
Denote $ \widetilde{\m{Y}}_{j,:} = \m{b}^j / d^{\cH}_j - \m{Y}_{j,:}$. Define the phase function $\phi\sbra{\m{z}} = \m{z}/\twon{\m{z}}$. Making the substitution $\m{Z}_{j,:} = \twon{\m{Z}_{j,:}} \phi\sbra{\m{Z}_{j,:}}$ yields that
\equ{
	\min_{\twon{\m{Z}_{j,:}},\phi\sbra{\m{Z}_{j,:}}} \sbra{\twon{\m{Z}_{j,:}}}^p + \mu d^{\cH}_j \twon{ \twon{\m{Z}_{j,:}} \phi\sbra{\m{Z}_{j,:}} - \widetilde{\m{Y}}_{j,:} }^2,
}
or equivalently,
\equ{
	\min_{\twon{\m{Z}_{j,:}},\phi\sbra{\m{Z}_{j,:}}} \sbra{\twon{\m{Z}_{j,:}}}^p + \mu d^{\cH}_j \twon{ \twon{\m{Z}_{j,:}}  - \widetilde{\m{Y}}_{j,:} \phi\sbra{\m{Z}_{j,:}}^H }^2.
}
Hence, the solution of $\phi\sbra{\m{Z}_{j,:}}$ is given by $\phi\sbra{\m{Z}_{j,:}} = \phi\sbra{\widetilde{\m{Y}}_{j,:}}$ and the solution of $\twon{\m{Z}_{j,:}}$ is given by
\equ{
	\twon{\m{Z}_{j,:}} \leftarrow  \text{prox}_{\abs{\cdot}^p,\beta_j}\sbra{ \norm{\widetilde{\m{Y}}_{j,:}}_2 }.
}
The final update is given by
\equ{
	\begin{split}
		\m{X}_{j,:}  \leftarrow 
		& \text{prox}_{\abs{\cdot}^p,\beta_j}\sbra{ \norm{\m{b}^j / d^{\cH}_j - \m{Y}_{j,:}}_2 } \\
		& \cdot \phi\sbra{\m{b}^j / d^{\cH}_j - \m{Y}_{j,:}} + \m{Y}_{j,:}.
	\end{split} \label{eq:row-impulsive}
}

When $g\sbra{\cdot} = \norm{\cP_{\Omega}\sbra{\cdot}}_{2,p}^p$, we have the update
\equ{
	\begin{split}
	 x_{jl} \leftarrow & \text{prox}_{\abs{\cdot}^p,\beta_j}\sbra{ \norm{\cP_{\Omega}\sbra{\m{b}^j / d^{\cH}_j - \m{Y}_{j,:}}}_2 } \\
	 & / \twon{ \cP_{\Omega}\sbra{\m{b}^j / d^{\cH}_j - \m{Y}_{j,:}} } \cdot \sbra{b^j_l / d^{\cH}_j - y_{jl}} + y_{jl}
	\end{split} \label{eq:row-impulsive_in}
}
if $\sbra{j,l}\in \Omega$ or \eqref{eq:out} otherwise.

\subsection{The ADMM Algorithm for Solving \eqref{eq:prb_T}} \label{append:T}
We introduce auxiliary variable $\m{Q} \in \bC^{(N+L)\times (N+L)}$ and get the augmented Lagrangian function:
\equ{
	\begin{split}
		\cL_{\text{T}} = g(\m{X}-\m{Y}) + \frac{\mu}{2} \frobn{\m{Q}-\begin{bmatrix}\m{Z} & \m{X}^H \\ \m{X} & \cT \m{t} \end{bmatrix} + \frac{1}{\mu}\m{\Lambda}}^2 - \frac{1}{2\mu} \frobn{\m{\Lambda}}^2,
	\end{split}	
}
where $\m{\Lambda}$ is the Lagrangian multiplier. 
The update for $\m{Q}$ is given by
\equ{
	\m{Q} \leftarrow \cP_{\bS_K^+}\sbra{\begin{bmatrix}\m{Z} & \m{X}^H \\ \m{X} & \cT \m{t} \end{bmatrix} - \frac{1}{\mu}\m{\Lambda}}.
}
Denote $\m{W} = \m{Q}+\frac{1}{\mu}\m{\Lambda}$ and write $\m{W}= \begin{bmatrix}\m{W}_{1} & \sbra{\m{W}_{2}}^H \\ \m{W}_{2} & \m{W}_{3} \end{bmatrix}$ as a block matrix as $\begin{bmatrix}\m{Z} & \m{X}^H \\ \m{X} & \cT \m{t} \end{bmatrix} $. 
The updates for $\lbra{\m{X},\m{Z},\m{t}}$ are given by
\equ{
	\begin{split}
		x_{jl} &\leftarrow y_{jl} + \text{prox}_{\abs{\cdot}^p,\mu}\sbra{ \sbra{\m{W}_2}_{jl} - y_{jl} }, \; \text{if} \; \lbra{j,l}\in \Omega, \\
		x_{jl} &\leftarrow \sbra{\m{W}_2}_{jl}, \; \text{if} \; \lbra{j,l} \notin \Omega, \\
		\m{Z} &\leftarrow \m{W}_1, \\
		\m{t} &\leftarrow (\cT^H\cT)^{-1} \cT^H \m{W}_3.
	\end{split}
}
The update for $\m{\Lambda}$ is given by
\equ{
	\m{\Lambda} \leftarrow \m{\Lambda} + \mu\sbra{\m{Q} - \begin{bmatrix}\m{Z} & \m{X}^H \\ \m{X} & \cT \m{t} \end{bmatrix} }.
}

\section*{Acknowledgement}
The authors would like to thank Prof. Petre Stoica for helpful comments on an earlier draft and the anonymous reviewers for their valuable comments that have improved the quality of the paper.

\bibliographystyle{IEEEtran}


\begin{IEEEbiography}
	[{\includegraphics[width=1in,height=1.25in,clip,keepaspectratio]{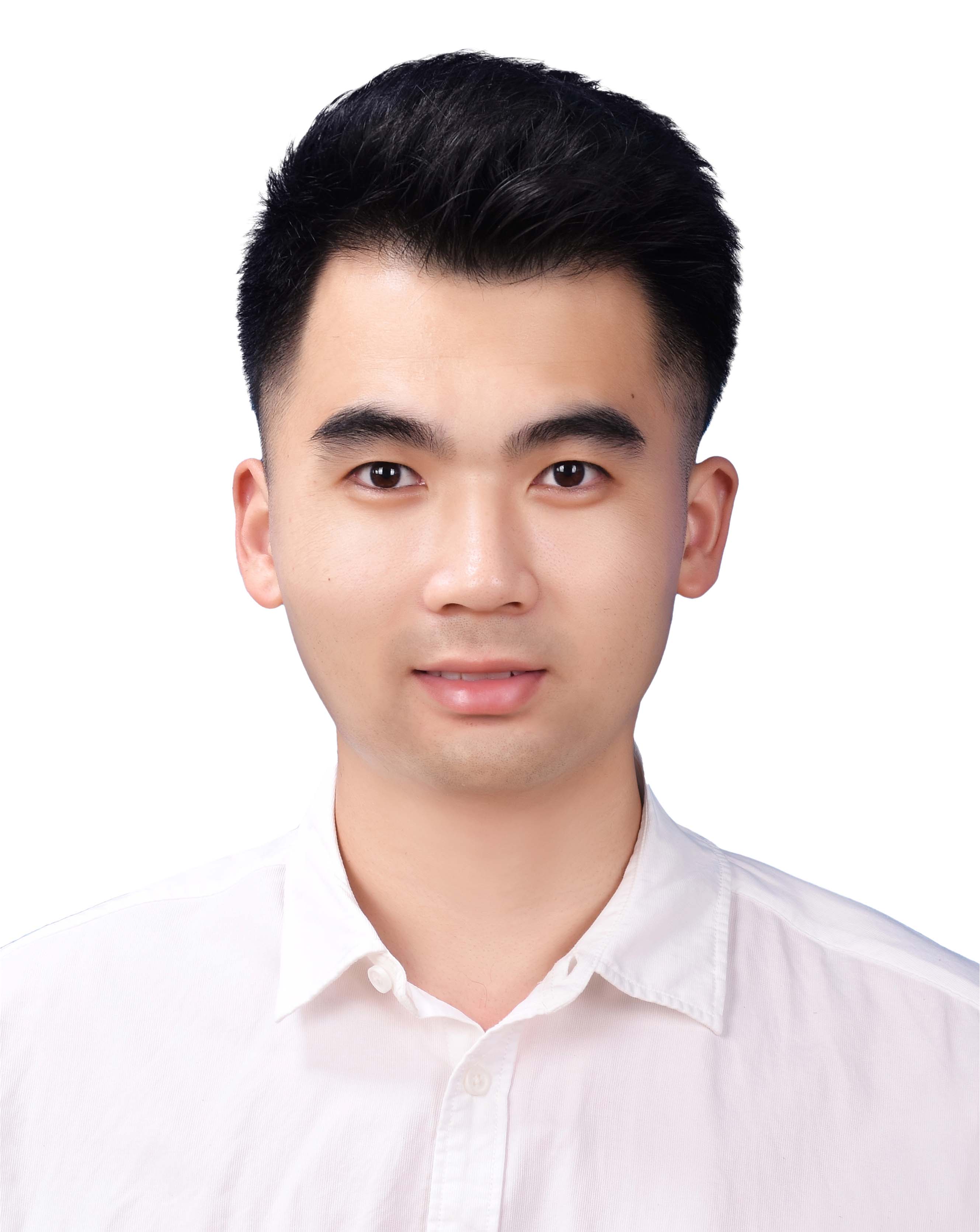}}]
	{Xunmeng Wu} received the B.S.~degree in mathematics from Xi'an Jiaotong University, China, in 2017. He is currently working toward the Ph.D.~degree in the school of mathematics and statistics of Xi'an Jiaotong University, China. His current research interests include compressed sensing, low-rank structured matrix recovery, and their applications in spectral analysis, array processing, and wireless communications.
\end{IEEEbiography}

\begin{IEEEbiography}
	[{\includegraphics[width=1in,height=1.25in,clip,keepaspectratio]{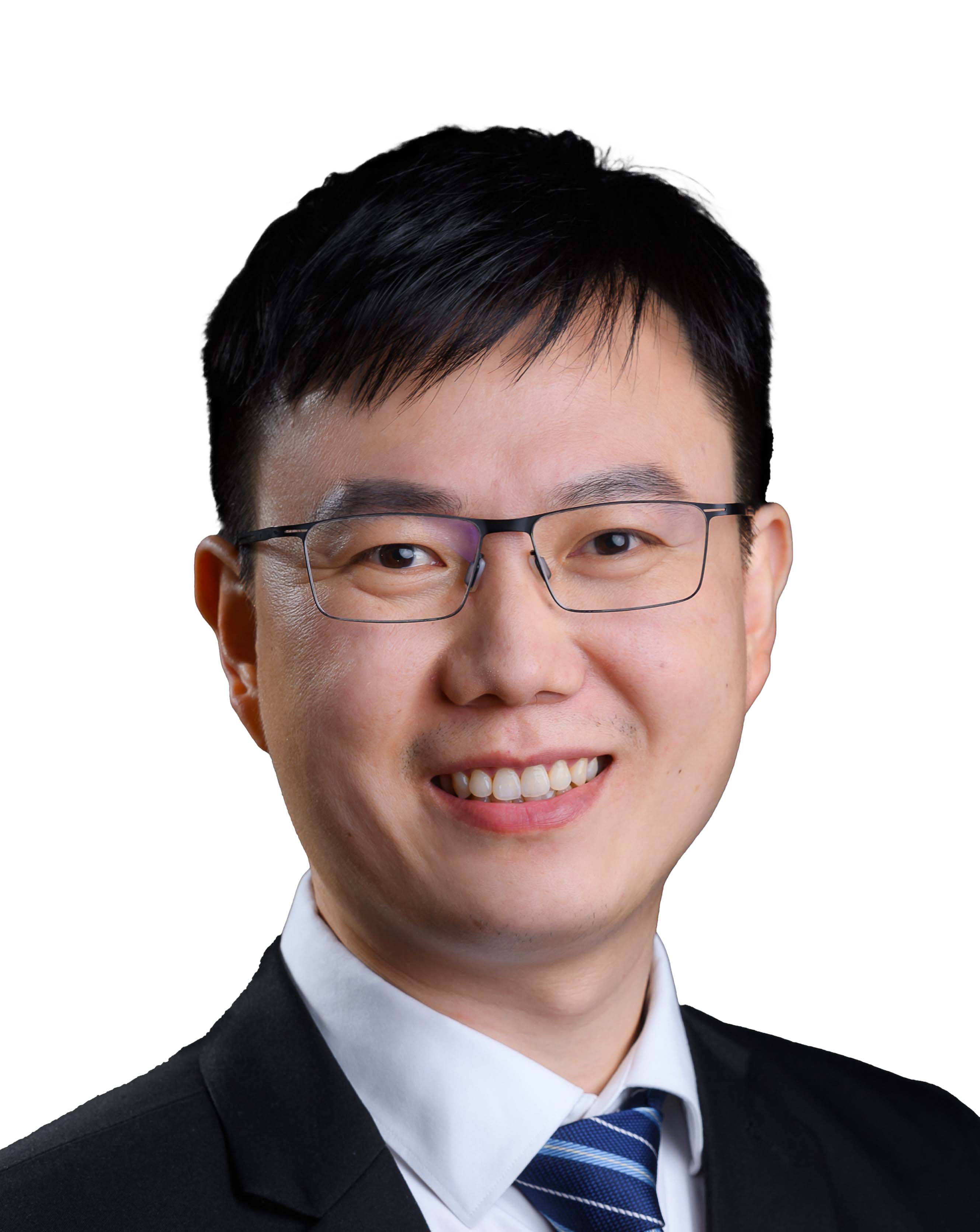}}]
	{Zai Yang} (S'11--M'14--SM'19) is a Professor of the School of Mathematics and Statistics, Xi’an Jiaotong University, China. He received the B.Sc.~degree in mathematics and M.Sc.~degree in applied mathematics from Sun Yat-sen (Zhongshan) University, China, in 2007 and 2009 respectively, and the Ph.D~degree in electrical and electronic engineering from Nanyang Technological University (NTU), Singapore, in 2014. He was a Research Associate and a Research Fellow of NTU from June 2013 to December 2015, and a Professor of the School of Automation, Nanjing University of Science and Technology (NJUST), China, from 2016 to 2018. His research interest is focused on mathematics of signal processing and wireless communications, especially on designing convex and nonconvex optimization algorithms with theoretical guarantees and developing mathematical tools for array signal processing, spectral analysis and wireless channel estimation. He was a leading tutorial presenter at EUSIPCO 2017. He is an IEEE senior member and a member of the Sensor Array and Multichannel (SAM) Technical Committee (TC) of the IEEE Signal Processing Society (2023-2025). He has been serving on the editorial board of {\em Signal Processing} (Elsevier) since 2017. He was awarded the NSFC Excellent Youth Science Foundation Grant in 2019.
\end{IEEEbiography}

\begin{IEEEbiography}
	[{\includegraphics[width=1in,height=1.25in,clip,keepaspectratio]{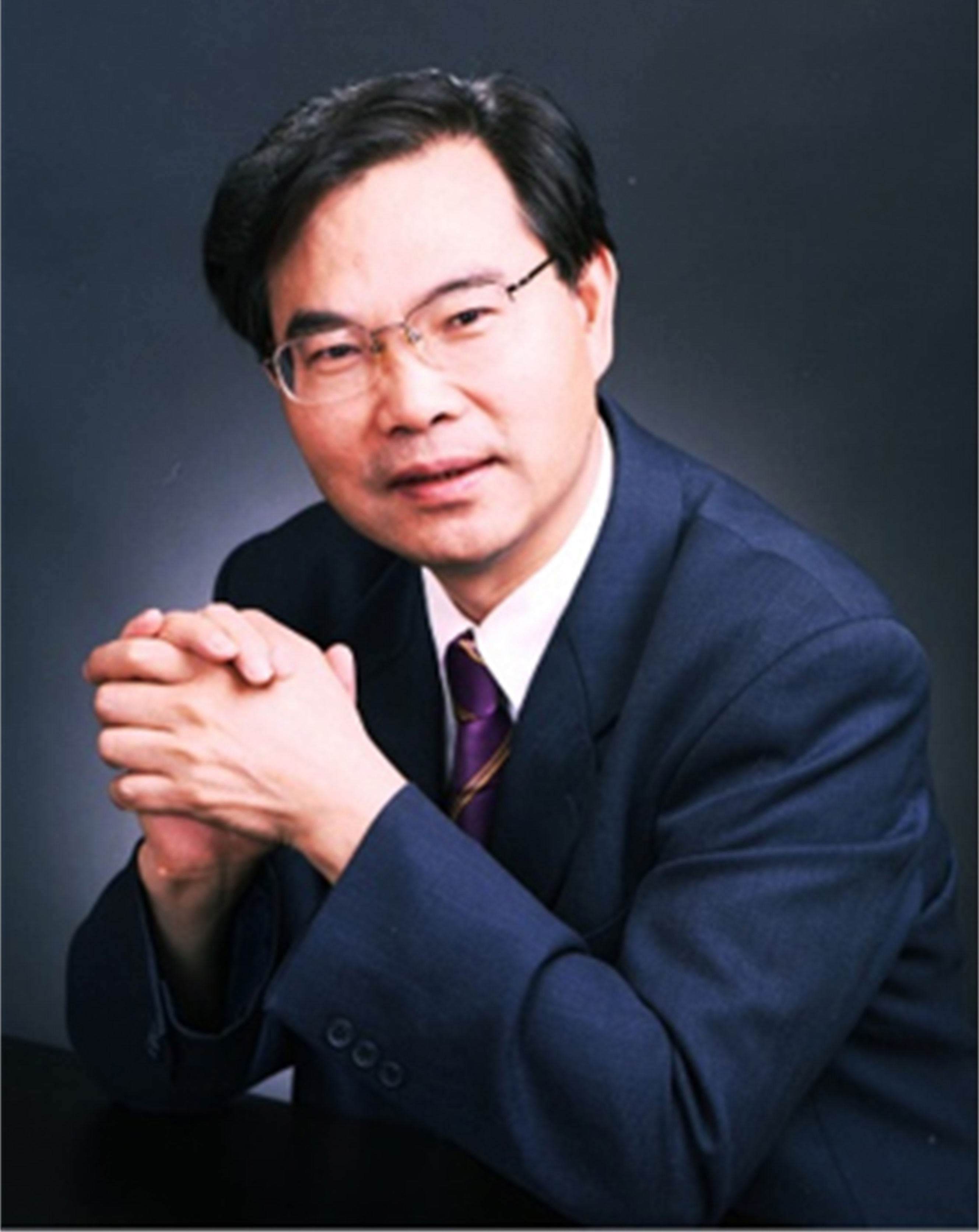}}]
	{Zongben Xu} received the PhD degree in mathematics from Xi’an Jiaotong University, Xi’an, China, in 1987. He currently serves as the Academician of the Chinese Academy of Sciences, the chief scientist of the National Basic Research Program of China (973 Project), and the director of the Institute for Information and System Sciences with Xi’an Jiaotong University. His current research interests include nonlinear functional analysis and intelligent information processing. He was a recipient of the National Natural Science Award of China, in 2007, and the winner of the CSIAM Su Buchin Applied Mathematics Prize, in 2008.
\end{IEEEbiography}

\end{document}